\documentclass[12pt,preprint]{aastex}

\def\deg{\ifmmode^\circ\else$^\circ$\fi}

\def\arcs{\ifmmode {^{\prime \prime}}\else $^{\prime \prime}$\fi}
\def\arcm{\ifmmode {^{\prime}}\else $^{\prime}$\fi}
\def\parcs{\sa=.07em \sb=.03em
     \ifmmode $\rlap{.}$^{\scriptscriptstyle\prime\kern -\sb\prime}$\kern -\sa$
     \else \rlap{.}$^{\scriptscriptstyle\prime\kern -\sb\prime}$\kern -\sa\fi}
\def\parcm{\sa=.08em \sb=.03em
     \ifmmode $\rlap{.}\kern\sa$^{\scriptscriptstyle\prime}$\kern-\sb$
     \else \rlap{.}\kern\sa$^{\scriptscriptstyle\prime}$\kern-\sb\fi}

\def\lya{{\rm Ly}$\alpha$}

\def\ha{\han}
\def\hb{\mbox{{\rm H}$\beta$}}

\def\h0{\mbox{H$_0$}}

\def\spose#1{\hbox to 0pt{#1\hss}}
\def\simlt{\mathrel{\spose{\lower 3pt\hbox{$\mathchar"218$}}
     \raise 2.0pt\hbox{$\mathchar"13C$}}}
\def\simgt{\mathrel{\spose{\lower 3pt\hbox{$\mathchar"218$}}
     \raise 2.0pt\hbox{$\mathchar"13E$}}}
\def\lsim{\rlap{$<$}{\lower 1.0ex\hbox{$\sim$}}}
\def\gsim{\rlap{$>$}{\lower 1.0ex\hbox{$\sim$}}}

\received{2002 July 30}
\begin{document}

\title{Emission Line Galaxies in the STIS Parallel Survey I:  Observations and Data Analysis\footnotemark}

\footnotetext[1]{Based on observations made with the NASA/ESA {\em
Hubble Space Telescope}, obtained from the data archive at the
Space Telescope Science Institute, which is operated by the
Association of Universities for Research in Astronomy, Inc., under
NASA contract NAS 5-26555.}

\author{
Harry I. Teplitz,\altaffilmark{2,}\altaffilmark{3} Nicholas R.
Collins,\altaffilmark{4} Jonathan P. Gardner, Robert S.
Hill,\altaffilmark{4} Sara R. Heap, Don J. Lindler,\altaffilmark{5} Jason Rhodes,\altaffilmark{6} 
\& Bruce E. Woodgate}

\affil{Laboratory for Astronomy and Solar Physics, Code 681,
Goddard Space Flight Center, Greenbelt MD 20771 \\Electronic mail:
hit@ipac.caltech.edu}

\altaffiltext{2}{Catholic University of America, Research Associate}

\altaffiltext{3}{New affiliation -- SIRTF Science Center}

\altaffiltext{4}{SSAI}

\altaffiltext{5}{Sigma Space Corporation}

\altaffiltext{6}{NRC Fellow}

\begin{abstract}

In the first three years of operation STIS obtained slitless spectra of
 $\sim 2500$\ fields in parallel to prime HST observations as part of the
 STIS Parallel Survey (SPS).  The archive contains $\sim 300$\ fields 
 at high galactic latitude ($|b|>30$) with  spectroscopic
exposure times greater than 3000 seconds.  
This sample contains 220 fields (excluding special
regions and requiring a consistent grating angle) observed between 
6 June 1997 and 21 September 2000, with
a total survey area of $\sim 160$\ square arcminutes.  At this depth, the
SPS detects an average of one emission line galaxy per three
fields.  We present the analysis of these data, and the identification
of 131 low to intermediate redshift
galaxies detected by optical emission lines.  The sample contains 78 objects with
emission lines that we infer to be redshifted [OII]3727 emission at $0.43<z<1.7$. 
The comoving number density of these objects is comparable to that of \ha-emitting
galaxies in the NICMOS parallel observations.
One quasar and three probable Seyfert galaxies are detected.  Many of the emission-line
objects show morphologies suggestive of mergers or interactions. The reduced data are
available upon request from the authors.

\end{abstract}


\section{Introduction}

The relative pointing of instruments on the Hubble Space
Telescope allows for parallel observation during targeted science
exposures.  Parallel fields are essentially random pointings on
the sky, being separated by 5-8 arcminutes from the prime target.
Extensive imaging of parallel fields was possible with WFPC2, and the
Medium Deep Survey obtained hundreds of fields (c.f. Griffiths et al.\ 1994).
With the addition of the slitless spectroscopic capabilities of STIS (Kimble et al.\ 1998) and
NICMOS (Thompson et al.\ 1998), parallel observations now allow HST to conduct redshift
surveys in large numbers of random fields without negatively
impacting the duty cycle of the telescope (see Gardner et al.\ 1998, 
McCarthy et al.\ 1999, Yan et al.\ 1999).  These untargeted
redshift surveys explore a new regime in galaxy evolution, as they
are less subject to a bias induced by cosmic variance than pencil beam
surveys (e.g. Cohen et al. 2000) and detect fainter objects than wide field
surveys (e.g. CFRS; Hammer et al.\ 1997).  The
slitless spectroscopy requires comparable deep imaging for object
identification; thus the redshift survey is accompanied by high
resolution imaging.

In this paper, we present the analysis of the deepest spectroscopic
fields in the STIS parallel survey (SPS).  STIS parallel spectra
were taken over the 5300-10000 \AA\ range during the course of the
program.  We have searched for emission line galaxies,
in particular [OII]-emitting
galaxies at redshift $0.43<z<1.7$.  Our motivation is to determine
the comoving density of star-formation at these redshifts as
inferred from [OII] emission (Teplitz et al.\ 2002; hereafter Paper II).  
We will discuss the target selection in section 2, the data reduction, line identification
and measurement in section 3, and the results in sections 4 and 5.

\section{Target Selection}

The SPS data were obtained in pairs of spectroscopic exposures and
direct images of each field.  The spectroscopic (dispersed) exposures
were taken with the G750L low resolution grating, which has a
dispersion of 4.92 \AA\ per pixel and a point source resolving power
of $\sim 750$\ at 7500 \AA.  The resolving power decreases for
extended objects in slitless spectroscopy, as the resolution element
is convolved with the galaxy profile. The spectra cover the range of
approximately 5300-10000 \AA, depending on object position within the
field.  The sensitivity falls off sharply at the blue end and
gradually at the red end.  We can detect [OII] at $0.43 < z < 1.68$,
[OIII]5007 at $z<1$, and H$\alpha$\ at $z< 0.52$.  The spectroscopic
exposures are binned by a factor of two in the spatial dimension in
order to improve read noise. The SPS includes G750L observations with
the grating angle set to provide different center wavelengths. To
ensure as uniform a dataset as possible, we limit our dataset to
spectra with a center wavelength of 7751\AA.  Direct imaging data are
taken without a filter, providing photometry integrated over
CCD-response limited wavelength range ($\sim 2200-11000$\ \AA, see
Baum et al.\ 1996), referred to as 50CCD magnitudes.

During the period 6 June 1997 and 21 September 2000, STIS observed
$\sim 2500$\ parallel fields.  Our goal is to study optical emission lines at intermediate redshift,
so our targets will be faint, and we select only
the deepest exposures.  Through trial and error we determined that 3000 seconds
were needed to reliably detect emission line objects in random fields.
We also require a deep image (300 seconds) to establish the 
wavelength scale, properly identify
the objects, and measure their photometric characteristics.  

STIS parallel fields lie 5 to 8 arcminutes from the targeted
position, depending on which instrument is prime.  Only those
fields that are not contaminated by the prime science target are
usable for this study.  Thus targets near the galactic plane
($|b|<30$), nearby galaxies, or galaxy clusters are excluded.  

The data were taken under a variety of observing plans.  In some cases all
imaging (or spectroscopic) exposures had the same integration time, in other
cases the integration times were optimized to fill the orbit.  The cosmic ray (CR) rejection 
algorithm is more effective in cases of equal integration time, and this should be 
considered in planning of future surveys. In a few fields the automated CR-rejection algorithm failed
to produce acceptable results, and we discard those fields from the survey.

After rejecting contaminated fields, 219 SPS fields 
meet our requirements. 
The sample contains fields with more than $30,000$\ seconds
of integration time, but the
distribution is strongly weighted towards the shortest allowed
exposure time (3000 seconds;  see Figure \ref{fig: exphist}).

Dithering of the observation was determined by the prime
instrument. Typically this provided a few tens of pixels in both
the $x$\ and $y$\ directions.  The regions on the edges with poor
overlap are not included in the survey.  Similarly, a bright star
or local galaxy can obscure significant area of the background
sky.  These obscured regions are not included in the survey area.
Table 1 lists the exposure time for both imaging and spectroscopic exposures for each
field, together with the pointing information and limiting flux
(see below), and the areal coverage of the good region.

We searched the NASA/IPAC Extragalactic Database (NED) for known objects in each 
field.  Three fields contain objects with previously reported redshifts, and many
fields contain previously reported objects without certain redshifts.  NED objects
are described in Table 2.

\section{Data Reduction}

The first step in the data reduction was identification of fields with
common pointing. Observations were considered as co-pointed if they
were centered within 5\arcs\ of each other, and they were processed as
a group.  Each exposure was processed through the basic data reduction
software, {\sc calstis}.  This software was developed by the STIS
Investigation Definition Team (IDT), and it is similar to the IRAF
version available from STScI.  Briefly, the data were bias and dark
subtracted, flat fielded, and bad pixels were repaired.  The bias
subtraction included an overscan fit to account for small temporal
variations.  Since the 2D spectroscopic exposures are binned by a
factor of two in $y$, the dark was binned in software before
subtraction.  We do not find that this binning introduces any
artifacts, but it is a step that must be checked carefully when
applied.  The flat fielding was performed in two steps: first, a
pixel-to-pixel flat was applied, and then a correction was made for
low frequency variation across the CCD induced by vignetting.  The
flat fields are monochromatic and so do not correct for fringing at
long wavelengths in the spectra. Fringing can cause increased noise in
the measurement of line fluxes at the $\sim 10$\%\ level. The bad
pixel repair was based on the best available (i.e., taken closest in
time) bad pixel mask from the archive.  Each bad pixel was replaced by
the average of its neighbors, and that pixel was flagged in the error
array.  For direct images the eight nearest pixels were used, but for
dispersed images only the left and right neighbors were averaged in
order to use data in the same spectrum.  A second bad pixel correction
was also performed on the output from {\sc calstis} in order to
``preclean" the data before CR rejection.  This second correction
checked for single, isolated pixels with outlying data values.

After the ``calstis" reduction, the offsets between dithered exposures
were calculated.  The offsets were determined by cross-correlation of
source positions in each image.  Spectroscopic exposures were assumed
to have the same offset as the image taken in the pair.  In some
cases, however, there were spectroscopic exposures that were taken
without a direct image at the exact position, and the shift needed to
be determined from the astrometry in the FITS header.  In these cases
the shifts are good to about a pixel, but are not as accurate as the
subpixel shifts determined from the extracted sources.  In order to
measure source positions in each image, a crude CR rejection must be
performed.  The algorithm looks for groups of high count-rate pixels
that are too compact or too highly peaked to be real objects.  Since
it is not necessary to preserve all real sources at this step, the
precleaning parameters are set to make the rejection severe. Sources
were identified using the SExtractor software (Bertin \& Arnouts
1996).

Once offsets had been determined, individual exposures were
registered, with bilinear interpolation for subpixel shifts.  
Cosmic rays in the images were identified in background
subtracted copies of the images.  The background was determined in
the direct images using two dimensional smoothing after source removal.
This procedure is modelled on  the SExtractor background
subtraction.  For the spectroscopic images, a one dimensional column-by-column
median was used for background subtraction.  The stack of registered images
was cleaned by omitting outliers in the distribution of pixel values separately at
each pixel position.  The cleaned images were combined into a final frame using the 
mean of pixels remaining at each position.  The reduced data were normalized to counts per 
second per unbinned pixel.

\subsection{Identifying, Extracting, and Measuring Emission Lines}

Each reduced spectroscopic frame was searched by eye for emission
lines.  Three of us (HIT, NRC, JR) examined every field.  Lines were
only considered real if there was unanimous agreement (though
independent discovery of each line was not required).  Figure
\ref{fig: obspair} shows an imaging/slitless spectrum pair, with
strong emission lines.  In many frames, bad pixels are present that
must be examined carefully by hand to distinguish them from real
emission lines.  At a minimum, a real line is required to have
significant flux in multiple pixels.  We set a significance threshold
of $\sigma=2.5$\ for the strongest emission line in an object in order
for it to be considered.  Further, we do not consider any possible
emission lines that cannot be associated with an object in the direct
image.  We detect emission lines in 134 objects.  Objects at low
redshift ($z<0.2$) are well resolved and the emission lines often
appear to originate in small knots; in those cases we did not measure
the line flux, but we will present the object identification and
redshift.

For extraction, the position of the spectrum in the 2D frame must be
known relative to the source position in the image.  The dispersed
frames have a small anamorphic magnification (less than 2\%), which is
included in the extraction.  Contemporaneous wavelength calibrations
cannot be taken for each individual parallel observation. The
wavelength scale is determined from an empirical fit to on-orbit data
taken early in the project. This scale appears to be stable at the
$\sim 1-2$\ pixel level, with variations arising from mechanism
non-repeatability and thermal motion.  The wavelength scale for an
extracted spectrum with pixels, $x_i$\ is

\begin{equation}
\lambda_i = \frac{x_i - c_0 - c_1 P_x}{d_0 + d_1 P_y} + \lambda_c
\end{equation} 
where ($P_x, P_y$) are the pixel position of the galaxy center in the
direct image, and the value of the constants depends on the grating
settings.  For the SPS, $c_0 = 4.50, c_1=0.99031, d_0=0.204737,
d_1=4.468\times 10^{-7},$\ and $\lambda_c = 7764.58$\ \AA.  We use a
linear wavelength scale, as the quadratic terms in the solution would
be at the subpixel level.  The value of ($P_x, P_y$) are accurate on
the integer pixel level.  Combined with the repeatability of the
scale, wavelength calibration uncertainty is $\sim 2-3$\ pixels.
Emission line wavelengths are measured to within 1 pixel, so the total
redshift uncertainty is on the order of 3 pixels ($\sigma_z \sim
0.004$\ for [OII] at $z=0.9$).

Spectra were extracted by simple summation along the columns covered
by the wavelength range.  The width of the extraction box was set by
hand for each object.  A small ($\simlt 15$\%) "slit correction" was
applied to the extraction based on the width of the box, compared to
the profile of the object.  Spectra were sky-subtracted based on two
sky windows, one on either side of the spectrum.  The width of these
windows, and their offset from the spectrum, were adjusted manually
for extraction, to achieve minimal contamination from other objects in
the field.  For each column in the spectrum, the average pixel value
in that column in each sky window is measured, and a line is fit to
the two points, giving a sky value at each pixel in the spectrum.
Before subtracting, these interpolated sky values are smoothed row by
row, using a median filter followed by two iterations of a mean
filter.  The extracted spectra were divided by the sensitivity
function to convert from DN/s/pixel to flux units.

Lines were measured in extracted spectra by interactively fitting a
gaussian to the line profile and integrating.  Continuum measurements
were made in the spectra themselves in regions near the line.  For
several bright objects the flux calibrated continuum measurements were
checked against the continuum value inferred from the clear aperture
photometry.  There is a trade off between the poor signal-to-noise in
the spectroscopic continuum, and the uncertainty in the slope of the
continuum needed to use the aperture photometry to infer the
continuum.  For consistency, the equivalent width (EW) of each line
was calculated from the local continuum measurement rather than the
photometric value.  This procedure leads to poor EW measurements in
the faintest, high EW objects. However, for the scientific analysis,
we will need to know the EW value with the greatest certainty for
objects with low EW values, and thus objects with better continuum
measurements (see Paper II).

\subsection{Single Line Objects}

 We detect only single lines in 94 of the 131 objects. In
order to assign a rest wavelength to single-line objects, it is
necessary to appeal to common sense constraints.  
We assume that any lone emission line is either \ha\ or the
next strongest optical line associated with star-formation, [OII]3727
(Kennicutt 1998).  \lya\ emission at $z > 5$\ is is not considered a
possibility, as there is often some flux detected blueward of the
line, and in most cases the objects are relatively bright.

The [OIII] doublet at 4959, 5007 \AA\ is also
sometimes strong in star-forming galaxies.  The ratio of the double is
1:3.1, so it is possible that a low signal to noise detection of a
single line could be 5007 (which would also imply weak \hb\ emission
at 4861\AA).  Typically, at low redshift [OIII] is undetected in
galaxies with EW(\ha)$<40$\ \AA, but can be 30\% the strength of \ha\ 
in the most vigorously star-forming objects (Kennicutt 1992).  In low
metallicity galaxies and HII regions, [OIII] can be as strong as \ha\,
or even stronger in some cases (Terlevich et al.\ 1991; Izotov, Thuan,
\& Lipovetsky 1994).  At high redshift ($z>2.5$), [OIII] has also been
detected galaxies with a line strength comparable to \ha\ in lower
metallicity ($1/3~Z_{\odot}$) galaxies (Pettini et al.\ 2001; Teplitz et al.
2000).  

However, in our spectra [OIII]5007 is not the most likely identification for a
single line.  At $z<0.5$, H-alpha would still be in the nominally
observable wavelength range, and at higher redshifts [OII] would be be
in the range at the blue end, and in some cases [OII] can be 50-200\% the strength of
[OIII] (e.g. Kobulnicky, Kennicutt, \& Pizagno 1999).  In addition, for the $\sim 10$\ single
line objects with good ($\sigma > 5$) detections, one might expect to
see some evidence of either [OIII]4959 or \hb\ if the line were [OIII]5007.

We do not necessarily expect to
see [NII] emission from the \ha\ emitters, as [NII] can be 0.1 to 0.3
the strength of \ha. Also, at low redshift [NII] may be hard to
resolve from \ha, given that spectra lose resolution for larger
objects. 

Given only the choice of [OII] and \ha, any line at wavelengths
shortward of 6563 \AA\ must be [OII].  Similarly an emission line
that would place an \ha-emitter at a low redshift ($z<0.2$) might
be expected to appear noticeably large and bright in the image.
We make this assumption, and thus do not consider the possibility
that the survey might detect extremely low mass galaxies at low redshift.

The ratio of \ha\ to [OIII]5007 emission is typically 2-3 to 1 (Kennicutt 1992) and
the ratio of \ha\ to \hb\ is 2.86 to one.
Thus a single line at the red end of the spectrum is taken to be
[OII] if there is no evidence of [OIII] or \hb\ where they would have been expected
had the line been \ha.

Finally, there are 28 objects that give no indication as to the
rest wavelength of the line.  Since the survey covers considerably
more volume for [OII] emitters, we assume that these lines are
[OII].  It should be kept in mind, then, that the density of
[OII] emitters inferred is an upper limit and that some downward
correction (25\%\ or less) might be expected.

The large percentage of single line objects in the sample is in part a result
of the limited spectral coverage.  Since objects near the edge of the field of
view have their spectral range truncated, [OII]-emitters for example are very 
likely to be single line objects.  Indeed, 40 of the 64 objects identified
as single line [OII] emitters are at redshifts that place the next strong
line ($\hb$) beyond the red end of the spectrum.  The observed wavelength range of
each spectrum is listed in Table 3.

A quality flag is assigned to the redshift of each galaxy.  A value of 4
indicates that there are multiple emission lines and the redshift
is secure.   A
value of 3 is assigned to galaxies for which there is a strong
reason for the assignment; that is one of the
following:\\
\begin{itemize}
\item (a) the wavelength of the single line is less than 6563\AA, or 

\item (b) the wavelength of the single line would place a faint
($m>23$) and/or compact object at $z<0.2$, or

\item (c) the redshift of the object if it were an \ha-emitter would
imply a non-detection of [OIII]5007.
\end{itemize}

A quality flag of 2 indicates that there is some potential problem
with the assignment.  Usually this indicates that the line is at a
blue wavelength (option $b$\ above) but that no [OIII]5007 is
detected at the red end of the spectrum.  
In a few cases, this non-detection is
in a region of poor spectroscopic sensitivity ($\lambda > 9000$\ \AA) and a quality flag of 1.5 is assigned
to differentiate it.  A non-detection of
[OIII] in an [OII] emitter is possible, depending on the
conditions in the galaxy.  The [OII] : [OIII] ratio can vary from 0.1 to 10.
These non-detections, if confirmed, would imply that
these galaxies had unusual metallicities or ionization parameters.
A quality flag of 1 indicates that the redshift is
poor, and that it is based solely  on the volume of the survey. 

\subsection{Measuring the Noise}

Given a weak or virtually nonexistent continuum detection (as in many SPS spectra), it is
hard to measure the signal to noise ratio of an emission line 
from the extracted spectrum.  Instead, the noise is measured in each
image and compared to the signal in the line.  Since each line is
a two dimensional pixel region before extraction, the noise can
measured in the sky in much the same manner as it would be for
photometry in imaging.  The specific method is to
randomly extract a box the size of the line at thousands of
positions across the frame.  The size is determined in the $y$\
direction by the width of the extraction box.  In the
$x$\ direction, the size is determined from the measured FWHM of the line.  Each
extraction includes a sky subtraction from neighboring sky
windows, just as in the spectroscopic extraction. The extracted values
should have a median of zero if there is no residual spatial
variation in the sky.  The standard deviation of the extractions
is the noise in the frame for a line with that spatial and
spectroscopic extent.  Table 1 lists the typical $1\sigma$\ limiting
fluxes for each field, based on an emission line from a galaxy with
$FWHM=0.39$\arcs, and a 7 pixel wide extraction box.  This size is 
slightly larger than the median for detected lines.

\section{Results}

We detected 131 emission line objects in the deep SPS fields.  The
direct image and extracted spectrum of each object are shown in figure
3-24.  Each object is described in Table 3.  Column (1) lists the name
of the object; columns (2) and (3) give the position of the object
based on the SExtractor pixel coordinates transformed to the World
Coordinate System, which for HST images is good to better than 1\arcs;
column (4) gives the redshift, and column (5) the redshift quality
flag; column (6) lists the photometric magnitude in the filterless
direct image; columns (7)-(9) give morphological parameters measured
by SExtractor; and column (10) lists the strongest lines detected in
the spectrum.  Line fluxes were only measured for those objects where
the integrated flux for the entire galaxy could be obtained; low
redshift objects with hotspot or diffuse emission were not measured,
and are flagged with redshift quality flag 0.  Table 4 gives the
emission line measurements.  Column (1) is the object name and column
(2) is the measured line; columns (3)-(5) give the line flux, its
signal-to-noise ratio, and the equivalent width; column (6) gives the
approximate continuum signal-to-noise in one resolution element near
the wavelength of the line.  The continuum is measured over many
resolution elements; however, the low signal-to-noise in the continuum
measurements demonstrates the large uncertainty in the EW
measurements.

\subsection{Notes On Individual Objects}


SPS J082344.12+292351.3  -- The [OIII]4959 line may be contaminated by a bad pixel.

SPS J095240.06+435834.9  -- This object shows three strong nuclei, suggesting either a triple nucleus or a projection 
of three edge on galaxies which appear ``stacked."  Emission lines are detected only from the nucleus
closest to the bottom of the image.

SPS J095628.10+694551.9  -- This bright spiral shows probable \ha\ emission from knots on the spiral arm.

SPS J100051.56+250858.6  -- There are two objects in the same row that could be associated with the emission line.  We assume
that the line is associated with the brighter one.

SPS J103640.31-034708.1  -- The blue end of the continuum (up to 6000 \AA) is contaminated by a bright object in the same row.

SPS J105700.03-034400.9  -- The object has a bright nucleus, perhaps indicating AGN, and a region of extended continuum suggesting
 a tail or jet.  The continuum is contaminated by bright galaxy on the same row.

SPS J105708.69-034310.8  -- There is a slight difference in inferred redshift from [OIII] and \ha.  They differ by 5\AA\ from the 
    prediction of the other.
    
SPS J105708.06-034320.1  -- This object has multiple nuclei in a ``train wreck" morphology. 

SPS J112057.83+232306.6  -- This object has a single emission line from the nucleus, although it has arm-like extensions from
which no emission is seen.

SPS J115623.08+550218.1  -- This spiral or irregular galaxy does not show a strong nucleus in the direct image, but has
very bright knots, at least 3 of which show strong \ha\ emission.

SPS J120113.92-185633.2  -- This object appears to have two nuclei, perhaps indicating that it is the result of an
interaction or merger, but only one of the nuclei shows strong line emission.  

SPS J123102.59+121424.8  -- This late-type spiral or irregular galaxy has many knots.  The nucleus shows clear \hb,[OIII], and \ha\ 
emission.

SPS J123348.78+023212.4  -- The object is very diffuse; nevertheless, the line looks definite on the dispersed image.

SPS J123703.08+140803.7  -- The objects appears to be the result of a merger or interaction of two objects.  They 
        are spatially separated along $x$.  

SPS J130015.23+275336.1  -- This spiral/irregular galaxy has many knots of bright continuum emission.  Some of the galaxy is 
outside the frame.  The nucleus and a few of the knots have a weak emission line, most likely \ha.

SPS J131810.43-000453.8  -- A point source with strong rest-frame UV emission lines, identified to be a quasar
at z=3.17.  

SPS J132511.10+301418.3  -- The continuum spectrum may be contaminated by other objects in the same row.

SPS J134429.63-000909.6  -- This object is a spiral galaxy with bright knots and also faint, extended continuum.  The emission
line from the bar looks double in spectrum, so it maybe emitted from two hotspots, though they are hard to 
match up with direct image.

SPS J134608.36+015922.6  -- This object shows an emission line from a double nucleus.  The line appears offset
in wavelength in the spectrum, because it is offset along the row.  The flux measurement 
is the sum of the flux from both nuclei.  

SPS J134721.93+021407.2 and SPS J134721.97+021405.7  -- A bright galaxy and a faint companion with a double
nucleus appear to be at the same redshift,
though perhaps the companions are simply knots in the outer regions of the galaxy itself.  The companion
shows relatively stronger oxygen emission than the galaxy.
 Also, the spacing between the nuclei is the same as the spacing between 5007 and 4959, so that the 5007 line from
the left nucleus coincides with 4959 from right nucleus; we measure 5007 from the right nucleus.

SPS J160956.65+653334.6  -- This bright galaxy shows strong emission lines including [FeVII]5721 and [OI]6300, at 
$z=0.157$.  The direct image of the source is not dominated by a bright nucleus, and we
infer that the object is a Seyfert.

SPS J162357.69+262830.2  -- This object has strong emission lines including [FeVI], and is likely a Seyfert 1.  The direct image
shows a strong nucleus, but the lines appear to originate mostly in the nebulosity offset from the nucleus
by $3-5$\ pixels.

SPS J162354.88+262803.5 -- This bright galaxy has a previously reported redshift of $z=0.47473$\ as a member of the
galaxy cluster MS $1621.5+2640$, which is centered at $z=0.4275$\ near this field, by the
Canadian Network for Observation Cosmology survey (CNOC; Ellingson et al.\ 1997).
The [OII] line that was likely the basis of the CNOC redshift 
is clearly detected in the SPS spectrum, but there the \hb\ and [OIII] doublet are absent (they would
not have been within the wavelength coverage for CNOC).  The CNOC redshift is supported by its membership
in the galaxy cluster, and thus it is likely that this galaxy is an example of a strong [OII] emitter which
we would have flagged as a problem (quality flag 2).  

SPS J162341.40+263433.0  -- The continuum may be contaminated by another object along $x$.  This object has previously been
reported by Ellingson et al.\ (1997) to be a likely member of the z=0.4275 cluster MS1621.5+2640.  There is 
a strong line in the spectrum that we interpret as \ha\ at z=0.369.  If the cluster redshift were accurate,
that line would be [OI]6300, but no [OIII] or \hb\ would be present in the spectrum.

SPS J164042.82+463843.1  -- This spiral galaxy shows stronger line emission from the lower arm than the nucleus.

SPS J171605.64+670411.9  -- The direct image of this galaxy shows strong continuum emission from a regular, S0 type morphology,
but connected to a fainter ``loop".  Likely \ha\ emission is seen both in the galaxy and faintly
in the loop.  The morphology is suggestive of an interaction.

SPS J201756.69-704702.1  -- This object is a low redshift spiral with many knots.  At least one knot has a clearly detectable
emission line, most likely \ha.

SPS J204757.09-194653.3  -- This object shows strong [OIII] emission but relatively weak Balmer emission, suggestive of an AGN.  The
direct image is only marginally resolved (FWHM = 2.8 pixels).

SPS J204757.58-194718.3  --  This huge spiral galaxy has at least 15 distinct knots in the direct image, most showing \ha\ emission
                    in the spectrum.  The inferred redshift of $z=0.111$\ is consistent within the uncertainty with
                    the previously reported value of $z=0.11526$\ (
                    De Carvalho et al.\ 1997).

SPS J224025.90-054736.3 -- This object appears to have three nuclei, with line emission clearly visible from all three
in the dispersed image.

SPS J224015.89-055449.7 -- This spiral galaxy has at least ten knots of continuum emission, some of which
also show clear \ha\ emission.

SPS J235158.92+243041.4 -- This spectacular spiral has many knots in the direct image, some of which show probable \ha\ emission.  

\subsection{Notes On Individual Fields}

J094317.0+465333.3  -- This field represents a surprising non-detection.  The field is located 
near the z=0.41 cluster Abell 0851, and contains a candidate cluster
member from the [OII] emission-line search conducted by Martin et al.\ (2000).  The wavelength
of redshifted [OII] at z=0.41 is not accessible to the SPS, but the candidate object shows no [OIII],\hb,
or \ha\ emission in the SPS spectrum.  The on-band flux excess of the Martin et al.\ candidate 
would have been $\sim 3\sigma$\ in the spectrum.   The morphology of the object is extremely irregular,
and the galaxy covers many pixels.  If the line emission is not primarily from a single hotspot it is 
likely that the emission lines are too faint to be detected in the spectrum.

J123630.1+621745.5  -- This field is one of the deepest STIS parallels.  Chen et al.\ (1999, 2000) searched
this field for high redshift Ly$\alpha$-emitters, but none were conclusively detected (see also
Stern et al.\ 2000).
We only detect two emission line objects at $z \simlt 1$.

\section{Discussion}

The deep SPS fields cover 160 square arcminutes (with 141 square
arcminutes of usable area) along 219 distinct lines of sight.  131
objects with emission lines were identified.  Forty-four fields have
multiple emission line objects.  Emission-line redshifts are inferred
ranging from z=0.034 to z=1.548.  Seventy-eight objects are identified
as having [OII] emission, of which 13 also have other emission lines.
Forty objects are identified primarily with \ha\ emission, and 13 with
[OIII] or \hb\ in the absence of \ha\ and [OII].  Of these, 37 have
secure redshifts (multiple lines; quality 4), 25 have quality 3, 28
have poor redshifts (quality 1), 15 have problem redshifts (quality
1.5 or 2), and 26 appear to have \ha\ emission for which we do not
measure line fluxes.

Figure \ref{fig: zhist}\ shows the distribution of inferred redshifts.
The detection of objects falls off at the
edges of the wavelength range as the sensitivity drops.  At the
lowest redshifts, the falloff in \ha\ objects is attributable to
the small volume covered by the survey.  Figure \ref{fig: maghist}\ shows
the distribution of apparent magnitude.

A rough calculation of the
density of detected objects can be made by considering the [OII]-emitters detected at 
the redshifts that place the [OII] line in the good regions of the sensitivity curve,
($0.5<z<1.2$).  The SPS surveys a comoving volume of $1.62\times 10^5$\ $h_{50}^{-3}$\ Mpc$^3$\ 
for $\Omega_M=1, \Omega_{\Lambda}=0$ (or $1.56\times 10^5$\ $h_{70}^{-3}$\ Mpc$^3$\ 
for $\Omega_M=0.3, \Omega_{\Lambda}=0.7$)
between those redshifts, and detects 63 [OII]-emitters,
giving a comoving number density of $3.7\times 10^{-4}$ $h_{50}^{-3}$\ Mpc$^{-3}$.  This density 
is slightly higher than that detected by the NICMOS parallels at $0.7<z<1.9$\ (McCarthy et al.\ 1999).
As McCarthy et al.\ (1999) point out, this density is lower than the density of galaxies 
brighter than $L^*$\ today (Ellis et al.\ 1996), not all of which are star-forming, 
but is comparable to that of Lyman Break Galaxies at $z\sim 3$\ (Steidel et al.\ 1996).  
The SPS sample of [OII] emitters is
sufficient to measure the comoving star-formation density at a median $z\sim 0.9$, and we will
present that analysis in Paper II.  

The SPS fields provide promising results for the detection of intermediate 
redshift galaxies in optical slitless spectroscopy of random fields.  While the survey is
subject to a number of biases (towards compact objects and away from low equivalent width lines),
it is an efficient means to detect distant, faint galaxies.  The high resolution of the direct images should also 
enable a study of the morphological properties of intermediate redshift galaxies.  The installation of the
Advanced Camera for Surveys (ACS; Ford et al.\ 2001) on HST will make a more extensive survey possible in the
coming years.  With its larger field of view and filtered imaging capabilities, ACS will refine
the procedures developed for the SPS program.

\acknowledgements

We thank the members of the Space Telescope Imaging Spectrograph
Investigation Definition Team (STIS IDT) for their encouragement of
this project. In particular, we acknowledge the contribution of Terry Beck, Ruth Bradley, 
Keith Feggans, Theodore R. Gull, 
Mary E. Kaiser, Philip C. Plait, 
Jennifer L. Sandoval, and Gerard M. Williger.
This research has made use of the NASA/IPAC Extragalactic Database (NED),which is operated
by the Jet Propulsion Laboratory, California Institute of Technology, under contract
with the National Aeronautics and Space Administration.

Funding for this publication was provided by NASA through Proposal number 
HST-AR-08380 submitted to the Space Telescope Science Institute, which is 
operated by the Association of Universities for Research in Astronomy, Inc., under
NASA contract NAS5-26555.  Support was also provided by  
the STIS IDT through the National Optical Astronomical Observatories
and by the Goddard Space Flight Center.  J.R.  was supported by the National Research
Council-GSFC Research Associateship.

\references

\reference{} Baum, S., et al.\ 1996, STIS Instrument Handbook, Version 1.0 (Baltimore, STScI)

\reference{} Becker, R. H., White, R. L., \& Helfand, D. J.\ 1995, ApJ, 450, 559

\reference{} Bertin, E. \& Arnouts, S. 1996, \aaps, 117, 393

\reference{} Chen, H.-W., Lanzetta, K. M., \& Pascarelle, S.\ 1999, Nature, 398, 586

\reference{} Chen, H.-W., Lanzetta, K. M., Pascarelle, S., \& Yahata, N.\ 2000, Nature, 408,562

\reference{} Cohen, J. G., Hogg, D. W., Blandford, R., Cowie, L. L., Hu, E., Songaila, A., Shopbell, P., \&
Richberg, K.\ 2000, ApJ, 538, 29

\reference{} Cristiani, S.,et al.\ 2000, A\&A, 359, 489

\reference{} De Carvalho, R. R., Ribeiro, A. L. B., Capelato, H. V., \& Zepf, S. E.\ 1997, ApJS, 110, 1

\reference{} Dressler, A., \& Gunn, J. E.\ 1992, ApJS, 78, 1

\reference{} Ellingson, E., Yee, H. K. C., Abraham, R. G., Morris, S. L.,
        Carlberg, R. G., \& Smecker-Hane, T. A.\ 1997, ApJS, 113,1

\reference{} Ellis, R. S., Colless, M., Broadhurst, T., Heyl, J., \& Glazebrook, K.\ 1996, MNRAS, 280,235

\reference{} Ford, H. C.; Advanced Camera for Surveys Science Team\ 2001, BAAS, 199.0802

\reference{} Gardner, J.P., et al.\ 1998, ApJ, 492, L99

\reference{} Giacconi, R., et al.\ 2002, ApJS, 139, 369         

\reference{} Godwin, Metcalfe, \& Peach\ 1983, MNRAS, 202, 113
                                                
\reference{} Griffiths, R.E., et al.\ 1994, ApJ, 435, L19

\reference{} Gruppioni, C., Zamorani, G., De Ruiter, H. R., Parma, P., Mignoli, M., \&
                                        Lari, C.\ 1997, MNRAS, 286, 470

\reference{} Hammer, F., et al.\ 1997, ApJ, 481, 49

\reference{} Hu, E. M., McMahon, R. G., \& Egami, E.\ 1996, ApJ, 459, L53

\reference{} Infante, L., Pritchet, C. J., \& Hertling, G.\ 1995, Journal of Astronomical Data, 1, 2

\reference{} Izotov, Y. I., Thuan, T. X., Lipovetsky, V. A.\ 1994, ApJ, 435, 647

\reference{} Kennicutt, R.C., Jr.\ 1992, ApJ, 338, 310

\reference{} Kennicutt, R. C., Jr.\ 1998, ARA\&A, 36, 189

\reference{} Kimble, R.A., et al.\ 1998, ApJ, 492, L83

\reference{} Kobulnicky, H.A., Kennicutt. R.C. Jr., \& Pizagno, J.L.\ 1999, ApJ 514, 544


\reference{}  Maddox, S. J., Sutherland, W. J., Efstathiou, G., \& Loveday, J.\ 
                          1990, MNRAS, 243, 692

\reference{}  Margoniner, V., \& De Carvalho, R.\ 2000, AJ, 119, 1562

\reference{} Martin, C., Lotz, J., \& Ferguson, H.\ 2000, ApJ, 543, 97

\reference{}  McCarthy, P.J., et al.\ 1999, ApJ,520, 548

\reference{}  Oort, M.\ 1987, A\&AS, 71,221

\reference{} Pettini, M., Shapley, A.E., Steidel, C.C., Cuby, J.-G., Dickinson, M., Moorwood, A.F.M., 
Adelberger, K.L., \& Giavalisco, M. 2001, ApJ, 554, 981

\reference{} Steidel, C.C., Giavalisco, M., Pettini, M., Dickinson, M.,
\& Adelberger, K.L., 1996, ApJ Letters 462, L17

\reference{} Stern, D., Eisenhardt, P., Spinrad, H., Dawson, S., van Breugel, W., Dey, A., de Vries, W., 
\& Stanford, S.A. 2000, Nature, 408, 560

\reference{}  Storrie-Lombardi, L. J., McMahon, R. G., Irwin, M. J., \& Hazard, C.\ 1996, ApJ, 468, 121

\reference{} Teplitz, H.I., et al.\ 2000,ApJ, 533, L65 

\reference{} Teplitz, H.I, Collins, N.R, Gardner, J.P., Hill, R.S., \& Rhodes, J.\ 2002 submitted 

\reference{} Terlevich, R., Melnick, J., Masegosa, J., Moles, M., Copetti, M. V. F.\ 1991, A\&AS, 91, 285

\reference{}  Thompson, R. I., Rieke, M., Schneider, G., Hines, D. C., \& Corbin, M. R.\ 1998, ApJ, 492, L95

\reference{} Williams, R.E., et al.\ 1996, AJ, 112, 1335

\reference{}  Yan, L., McCarthy, P.J., Freudling, W., Teplitz, H.I., Malumuth, E.M., 
\& Weymann, R.J., Malkan, M.A.\ 1999, ApJ, 519, L47

\clearpage


\begin{deluxetable}{llrrrr}
\tabletypesize{\scriptsize}
\tablecaption{Deep SPS Fields \label{tbl-1}}
\tablewidth{0pc}
\tablehead{
\colhead{$\alpha$(2000)} & 
\colhead{$\delta$(2000)}   & 
\colhead{Img. Exptime}   &
\colhead{Spc. Exptime} &
\colhead{1$\sigma$\ flux\tablenotemark{1}}  & 
\colhead{area} \\
\colhead{} & 
\colhead{} & 
\colhead{(s)} & 
\colhead{(s)} & 
\colhead{} & 
\colhead{(sq. arcmin.)} 
}
\startdata
 00 16  2.0  & +15 47  9.5  &          600 &         3638 &            5 & 0.63 \\ 
 00 16 27.9  & +16 01 46.9  &          600 &         3631 &            5 & 0.43 \\ 
 00 17 17.1  & +16 02 20.8  &          600 &         3630 &            4 & 0.64 \\ 
 00 17 26.8  & +15 48  8.2  &          600 &         3632 &            5 & 0.64 \\ 
 00 37 25.7  & +09 07 28.9  &         1350 &         5239 &           19 & 0.55 \\ 
 00 38  4.3  & +13 18 11.8  &         1700 &         4284 &            5 & 0.58 \\ 
 01 00 31.9  & -33 28 21.6  &         1200 &         7200 &            4 & 0.67 \\ 
 01 03 29.8  & +13 15 15.3  &         7296 &         4953 &            4 & 0.64 \\
\enddata

\tablenotetext{1}{($10^{-17}$ erg cm$^{-2}$ s$^{-1}$)}

\end{deluxetable}


\clearpage


\begin{deluxetable}{lllr}
\tabletypesize{\scriptsize}
\tablecaption{NED Objects in the Deep SPS Fields \label{tbl-2}}
\tablewidth{0pc}
\tablehead{
\colhead{$\alpha$(2000)} & 
\colhead{$\delta$(2000)}   & 
\colhead{NED Objects} &
\colhead{ref. \tablenotemark{1}} 
}
\startdata
     03 12 40.6  & -55 25 36.9  &   APMUKS(BJ) B031118.74-553640.9                           &  1  \\ 
     03 13 53.8  & -55 01 14.1  &    [GZd97] 1.4GHz 20 \& 21                                 &  2  \\ 
     03 14  1.8  & -55 25  8.4  &    APMUKS(BJ) B031242.66-553605.6                          &  1 \\
     03 14 40.6  & -55 06 17.4  &     APMUKS(BJ) B031319.21-551732.9                         & 1 \\ 
     03 32  6.0  & -27 44 15.4  &    CDF-S\tablenotemark{2};  EIS U21; z=3.462               &  3  \\ 
     08 48 38.2  & +44 48 37.5  &      65W 106                                                &  4  \\
     08 49 10.7  & +37 26 12.1  &     FIRST J084910.4+37263                                  &  5  \\ 
     09 43 16.9  & +46 56  3.3  &    ABELL 0851:[DG92] 044                                      &  6  \\
\enddata

\tablenotetext{1}{
1 -- Maddox et al.\ (1990); 2 -- Gruppioni et al.\ (1997); 3 -- Cristiani et al.\ (2000); \\
4 -- Oort (1987);  5 -- Becker et al.\ (1995); 6 -- Dressler \& Gunn (1992); \\
7 -- Martin et al.\ (2000); 8 -- Storrie-Lombardi et al.\ (1996); 9 -- Godwin et al.\ (1983); \\
10 -- Infante et al.\ (1995); 11 -- Margoniner \& De Carvalho (2000); 12 -- Ellingson et al.\ (1997); \\
13 -- De Carvalho et al.\ (1997)
}
\tablenotetext{2}{This field is a few arcminutes from the center of the Chandra Deep Field South (Giacconi et al.\ 2002).}

\tablenotetext{3}{This field was included in the Hu et al.\ (1996) Lyman-$\alpha$\ search for quasar companions.}

\tablenotetext{4}{This field, parallel to the Hubble Deep Field North (Williams et al.\ 1996) was analyzed independently
by Chen et al.\ (1999,2000) and Stern et al.\ (2000).}

\tablenotetext{5}{This field contains:  APMUKS(BJ) B130842.46-010229.8, B130844.25-010243.7 (z=0.209498), 
[MD2000] J131117.984-011812.47, [MD2000] J131118.783-011843.00, [MD2000] J131119.214-011830.18}
\end{deluxetable}


\clearpage


\def\ha{H$\alpha$}
\def\hb{H$\beta$}
\def\hg{H$\gamma$}
\def\hd{H$\delta$}

\begin{deluxetable}{lllrrrrrrll}
\rotate
\tabletypesize{\scriptsize}
\tablecaption{Detected Objects \label{tbl-3}}
\tablewidth{0pt}
\tablehead{
\colhead{obj name} &
\colhead{$\alpha$(2000)} & 
\colhead{$\delta$(2000)}   & 
\colhead{Redshift\tablenotemark{1}}   &
\colhead{z flag\tablenotemark{2}} &
\colhead{50CCD\tablenotemark{3}} &
\colhead{FWHM}  & 
\colhead{a/b} & 
\colhead{P.A.}&
\colhead{$\lambda$\ range}&
\colhead{Strong Lines} \\
\colhead{} &
\colhead{} & 
\colhead{} & 
\colhead{} & 
\colhead{} & 
\colhead{} &
\colhead{(arcsec)} & 
\colhead{} &
\colhead{(deg.)} &
\colhead{(\AA)}&
\colhead{Detected}  
}
\startdata
SPS J001716.56+160239.5 & 00:17:16.56 &  16:02:39.5 & 1.074 & 1 & 25.03 $\pm$\ 0.09 & 0.45 & 1.0 & -85                  & 6343-10000 & [OII] \\ 
SPS J010030.43-332823.4 & 01:00:30.43 & -33:28:23.4 & 0.702 & 3 & 24.61 $\pm$\ 0.05 & 1.34 & 1.7 & -40                  & 5300-8467 &  [OII] \\ 
SPS J010329.80+131508.9 & 01:03:29.80 &  13:15:08.9 & 0.819 & 4 & 22.64 $\pm$\ 0.03 & 0.33 & 1.2 & -27                  & 5392-10000 & [OII]-\hd-\hg-\hb-[OIII] \\ 
SPS J021048.23-393308.0 & 02:10:48.23 & -39:33:08.0 & 0.593 & 2 & 26.17 $\pm$\ 0.08 & 0.38 & 1.2 & 37                   & 5300-8598 &  [OII] \\ 
SPS J022828.08-011434.0 & 02:28:28.08 & -01:14:34.0 & 0.151 & 0 & 21.14 $\pm$\ 0.03 & 2.79 & 1.8 & 45                   & 7123-10000 & \ha \\ 
SPS J030221.47+001150.6\tablenotemark{4} & 03:02:21.47 &  00:11:50.6 & 0.920 & 3 & 24.41 $\pm$\ 0.09 & 0.97 & 1.9 & -19 & 5300-9180 &  [OII] \\ 
SPS J030210.18+001140.0\tablenotemark{4} & 03:02:10.18 &  00:11:40.0 & 1.092 & 1 & 25.20 $\pm$\ 0.13 & 0.45 & 1.3 & 50  & 5375-10000 & [OII] \\ 
SPS J031357.20-550113.5 & 03:13:57.20 & -55:01:13.5 & 0.688 & 3 & 23.13 $\pm$\ 0.03 & 0.72 & 1.3 & 62                   & 5300-8445 &  [OII] \\
\enddata

\tablenotetext{1}{The redshift uncertainty is $\sigma_z \sim 0.004$.}

\tablenotetext{2}{Redshift flag: \\
0   -- flux is in small knots, so we do not measure an integrated flux for lines\\
1   -- single line \\
1.5 -- single line, inferred to be [OII], but with redshifted \hb,[OIII] falling in region of low sensitivity \\
2   -- single line, inferred to be [OII], but with redshifted \hb,[OIII] falling in region of good sensitivity \\ 
3   -- single line with good reason to infer line (see text)
4   -- multiple lines}                              

\tablenotetext{3}{Photometry for the filterless direct image, given on the STMAG system.}

\tablenotetext{4}{ SPS J030221.47+001150.6 is detected in the J030220.8+001153.9 field, while SPS J030210.18+001140.0
is detected in the J030210.9+001155.3 field.}

\tablenotetext{5}{ SPS J095615.40+694535.3 is detected in the J095618.3+694537.0 field, while SPS J095628.10+694551.9 is detected 
in the J095623.8+694540.6 field.
}

\tablenotetext{6}{ This object is detected in both the J1213008+121420.7 and J1213013+121418.3 fields.}

\end{deluxetable}


\clearpage


\def\Halpha{H$\alpha$}
\def\Hbeta{H$\beta$}
\def\Hgamma{H$\gamma$}
\def\Hdelta{H$\delta$}

\begin{deluxetable}{llrrrr}
\tabletypesize{\scriptsize}
\tablecaption{Detected Objects \label{tbl-4}}
\tablewidth{0pt}
\tablehead{
\colhead{obj name} &
\colhead{line} & 
\colhead{line flux\tablenotemark{1}}   & 
\colhead{$N_{\sigma}$}   &
\colhead{$W_{obs}$} &
\colhead{continuum $\sigma$\tablenotemark{2}} \\
\colhead{} &
\colhead{} & 
\colhead{} & 
\colhead{} & 
\colhead{(\AA)} & 
\colhead{} 
}
\startdata
SPS J001716.56+160239.5 & [OII] & 1.35 & 3.7 & 141 & 0.5 \\ 
SPS J010030.43-332823.4 & [OII] & 2.34 & 5.1 & 219 & 0.5 \\ 
SPS J010329.80+131508.9 & [OII] & 4.74 & 10.5 & 156 & 1.5 \\ 
        & \Hdelta & 0.250 & 2.1 & 7 & 2.9 \\ 
        & \Hgamma & 0.860 & 3.2 & 30 & 1.6 \\ 
        & \Hbeta & 1.20 & 3.3 & 58 & 0.7 \\ 
        & [OIII]4959 & 5.94 & 8.0 & 240 & 0.8 \\ 
        & [OIII]5007 & 12.6 & 14.5 & 481 & 0.7 \\
\enddata

\tablenotetext{1}{($10^{-16}$\ erg cm$^{-2}$\ s$^{-1}$)}
\tablenotetext{2}{Continuum signal to noise ratio in a two pixel resolution element.}

\end{deluxetable}


\clearpage

\begin{figure}[ht]
\plotone{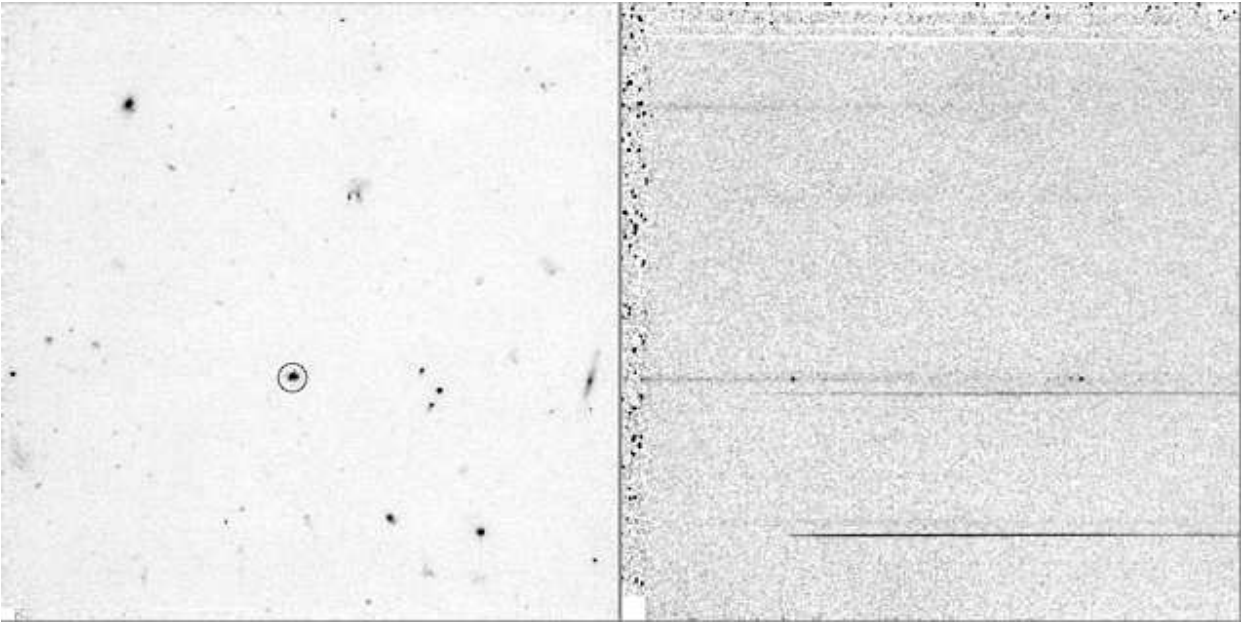}
\caption{An image/spectrum pair of the J010329+131515 field from the deep SPS, containing   
one of the highest signal-to-noise ratio emission line objects (circled
in the direct image); [OII],
the [OIII] doublet, and \hb\ are clearly visible in the 2 dimensional spectrum.   The poor cosmic ray rejection is 
visible in the non-overlap regions resulting from telescope offsets.
  \label{fig: obspair}}
\end{figure}
\clearpage

\begin{figure}
\plotone{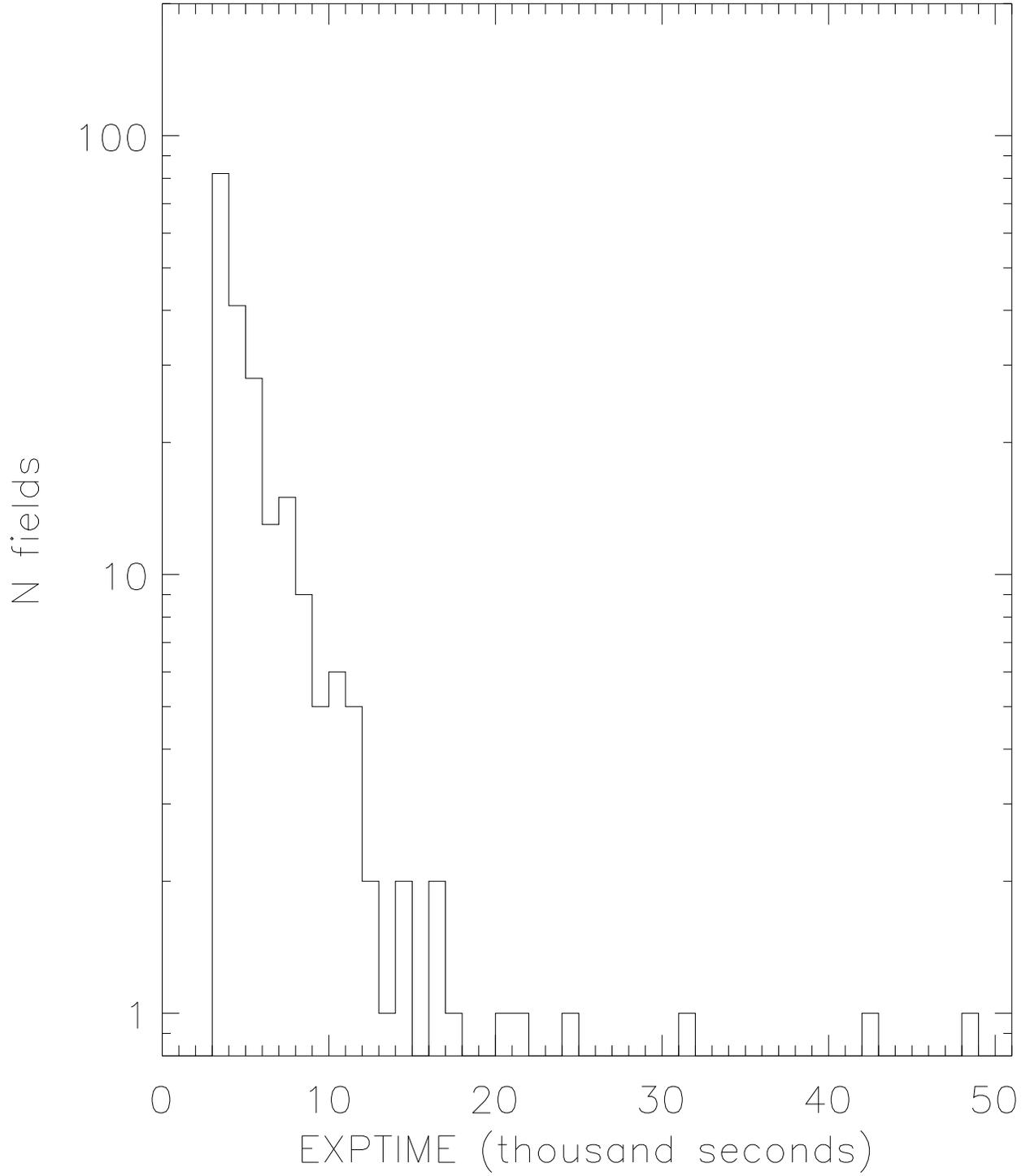}
\caption{The distribution of exposure times in the deep SPS fields meeting the selection criteria (see text)  
\label{fig: exphist} }
\end{figure}
\clearpage

\begin{figure}
\vskip -1in
\plotone{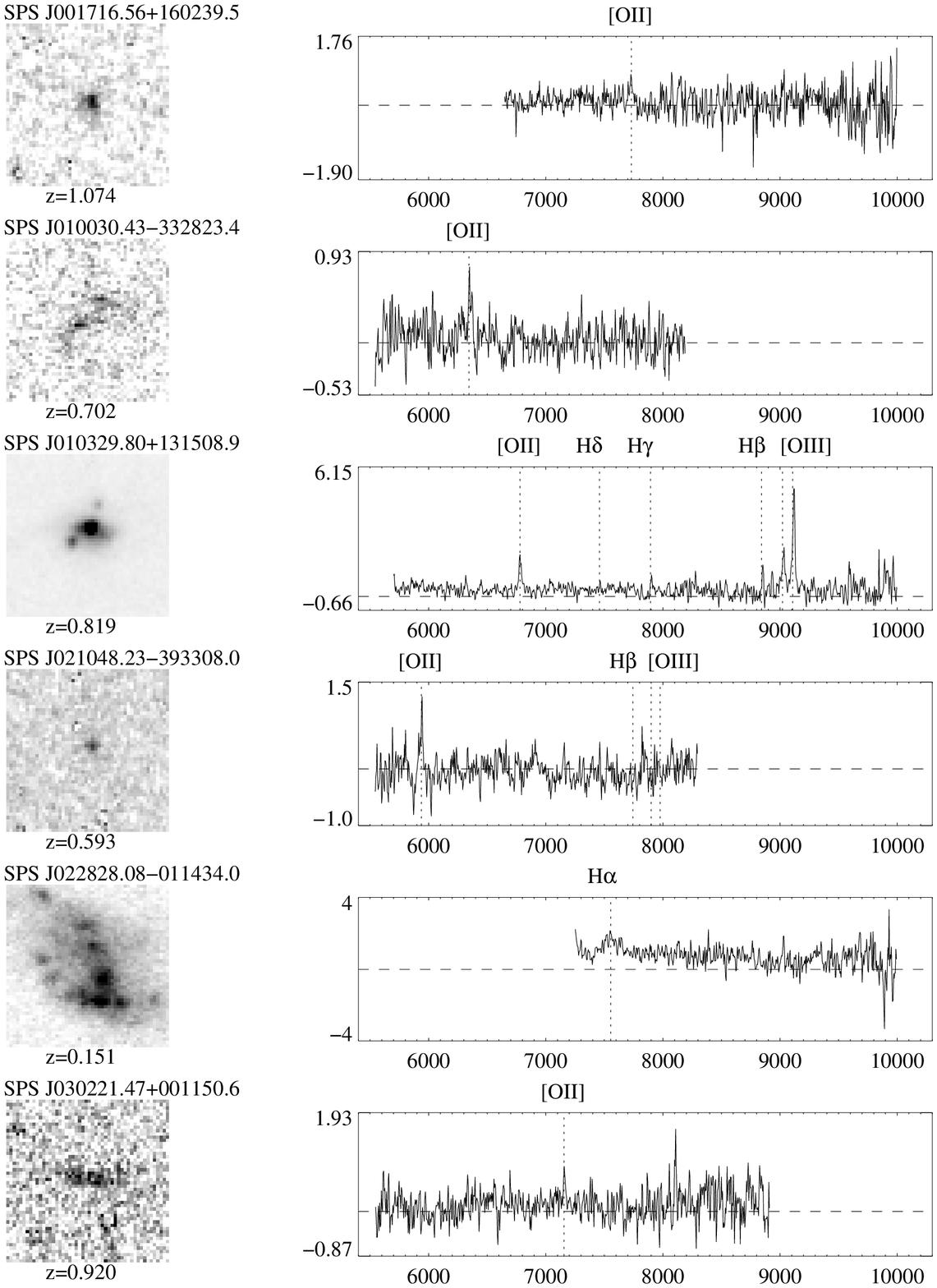}
\vskip -0.5in
\caption{The direct image and extracted spectrum for SPS emission line objects.  The postage
stamp images are 2.5\arcs\ on each side, with the orientation of the original data.  The postage 
stamp is centered on the region of strongest line emission, which may not coincide with the
center of the object.  The position
of the \ha,\hb,[OIII], and [OII] emission lines are indicated by dotted lines in each spectrum whether they
are detected or not.  The wavelength of other emission lines is indicated if they are measured and 
presented in Table 3.  \label{fig: postagestamp} }

\end{figure}
\clearpage

\begin{figure}
\epsscale{0.90}
\plotone{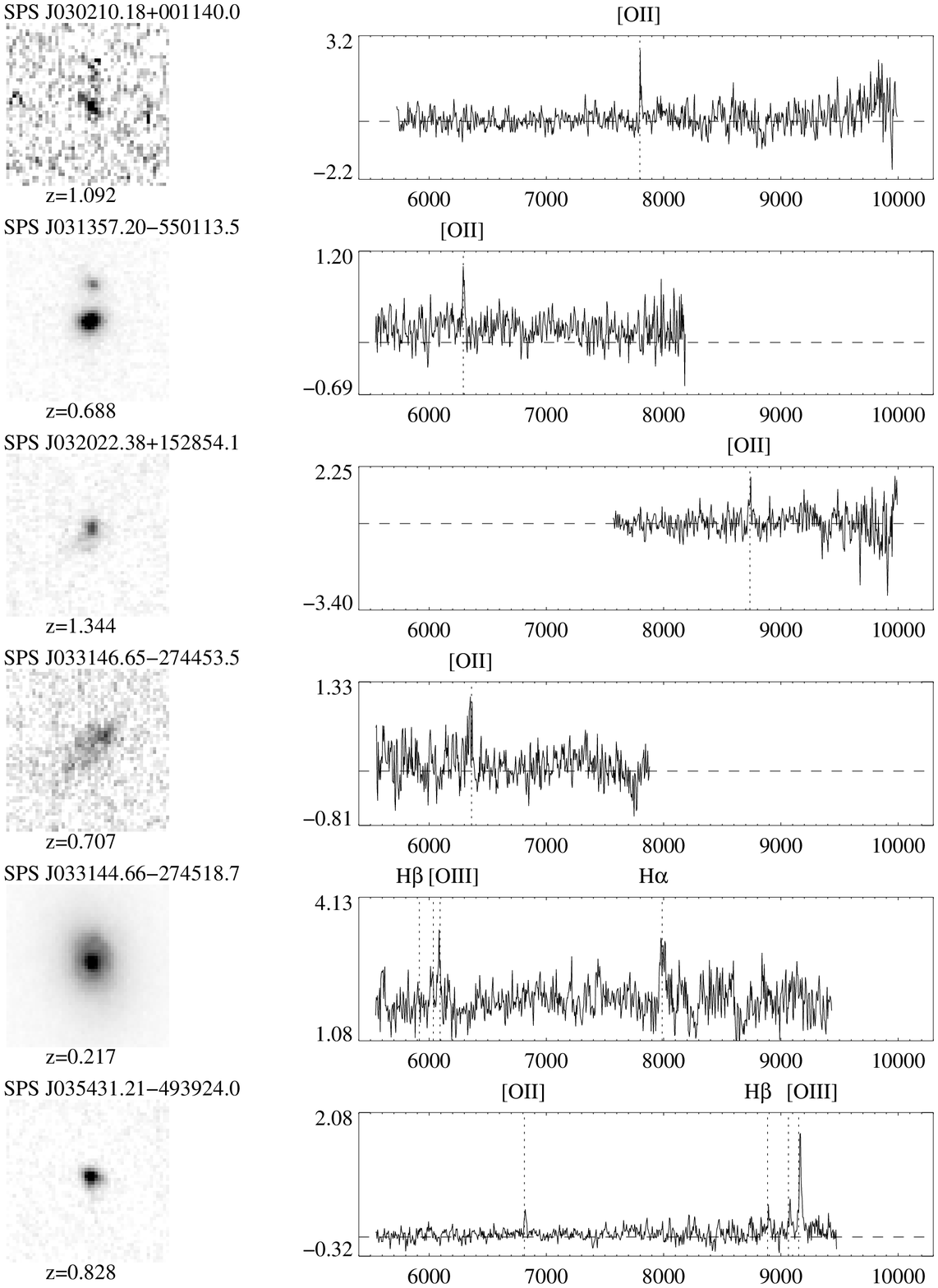}
\caption{The direct image and extracted spectrum for SPS emission line objects, as in Figure 3.}
\end{figure}

\clearpage
\begin{figure}
\plotone{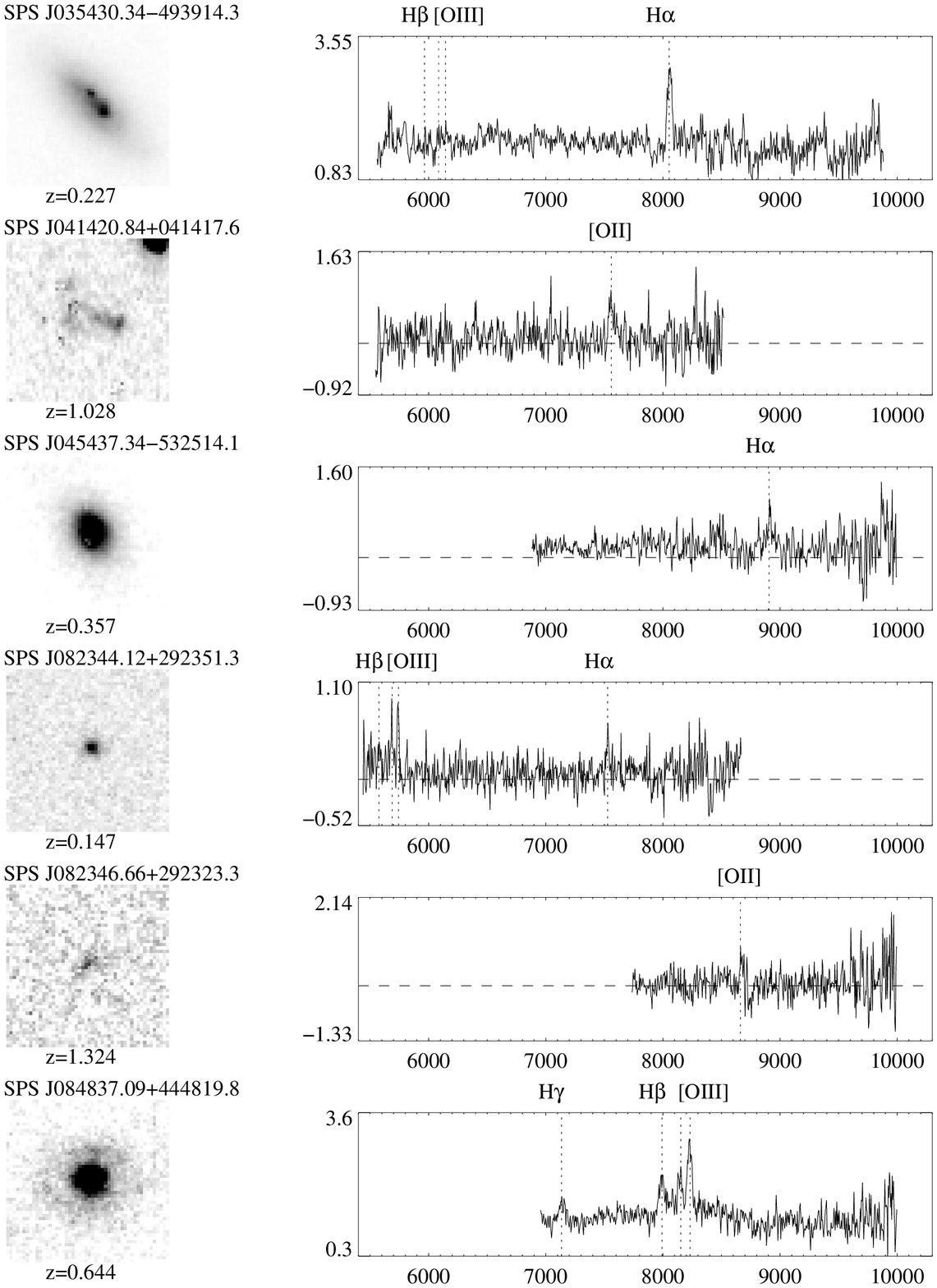}
\caption{The direct image and extracted spectrum for SPS emission line objects, as in Figure 3.}
\end{figure}

\clearpage
\begin{figure}
\plotone{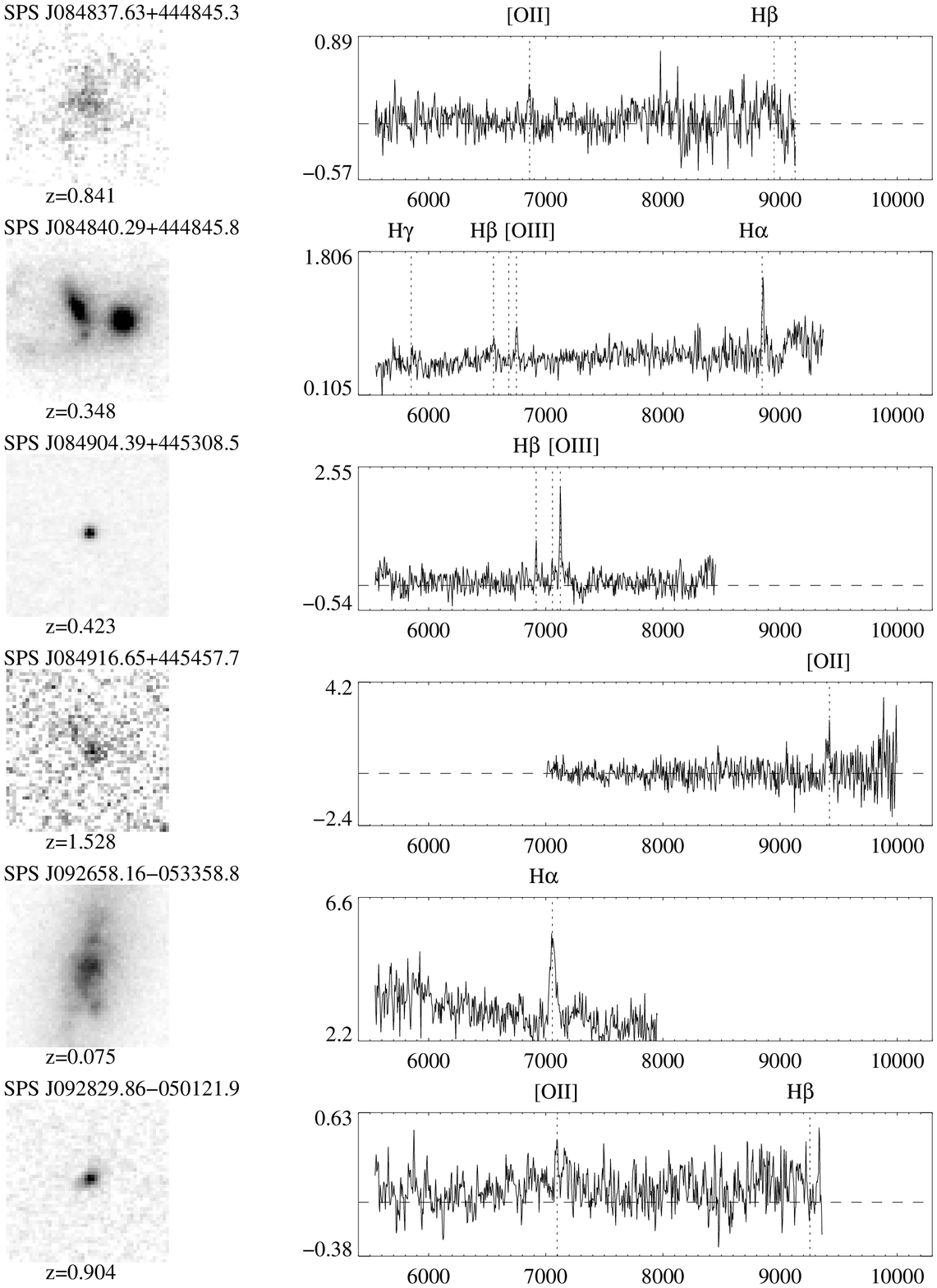}
\caption{The direct image and extracted spectrum for SPS emission line objects, as in Figure 3.}
\end{figure}

\clearpage
\begin{figure}
\plotone{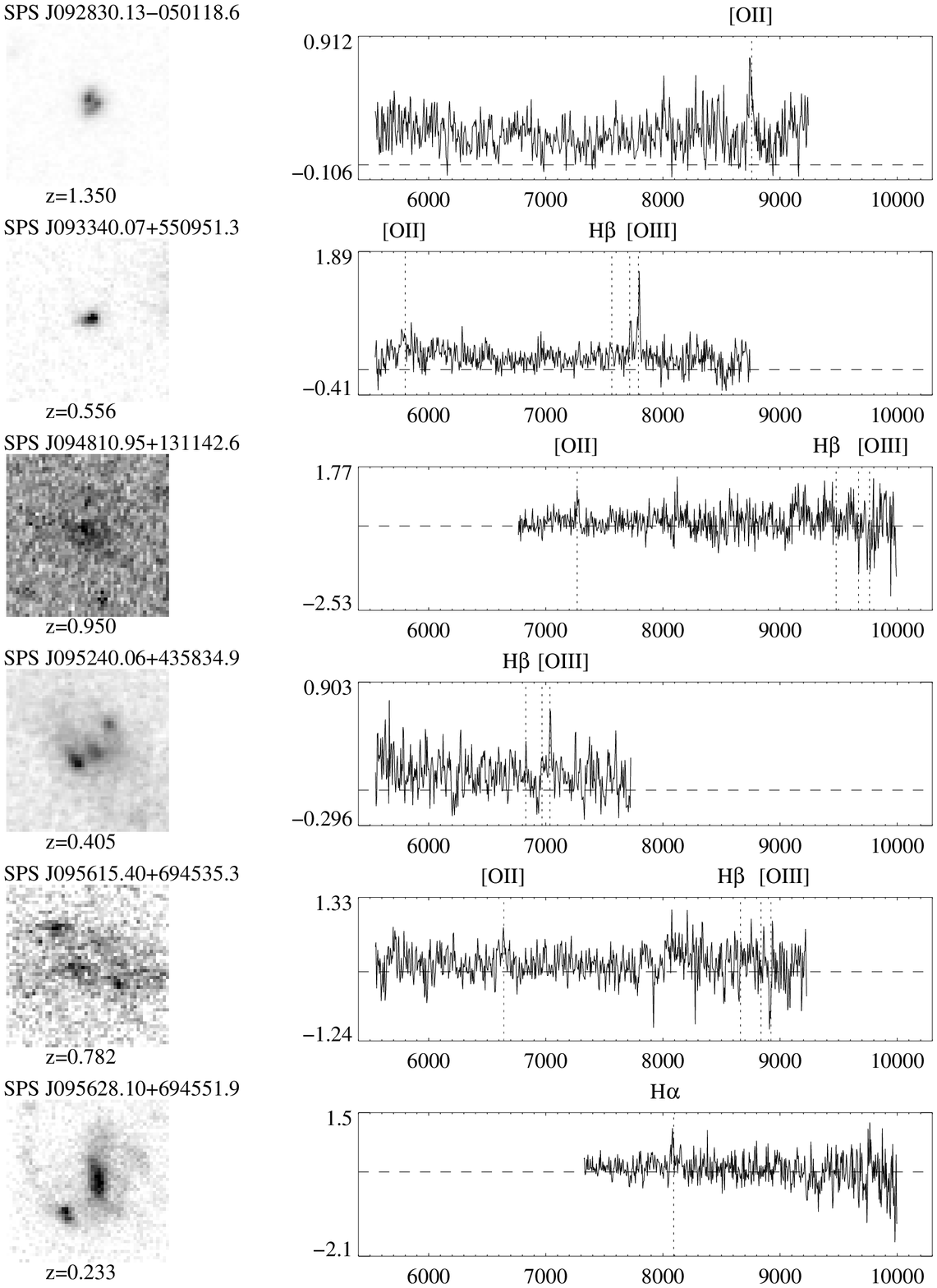}
\caption{The direct image and extracted spectrum for SPS emission line objects, as in Figure 3.}
\end{figure}

\clearpage
\begin{figure}
\plotone{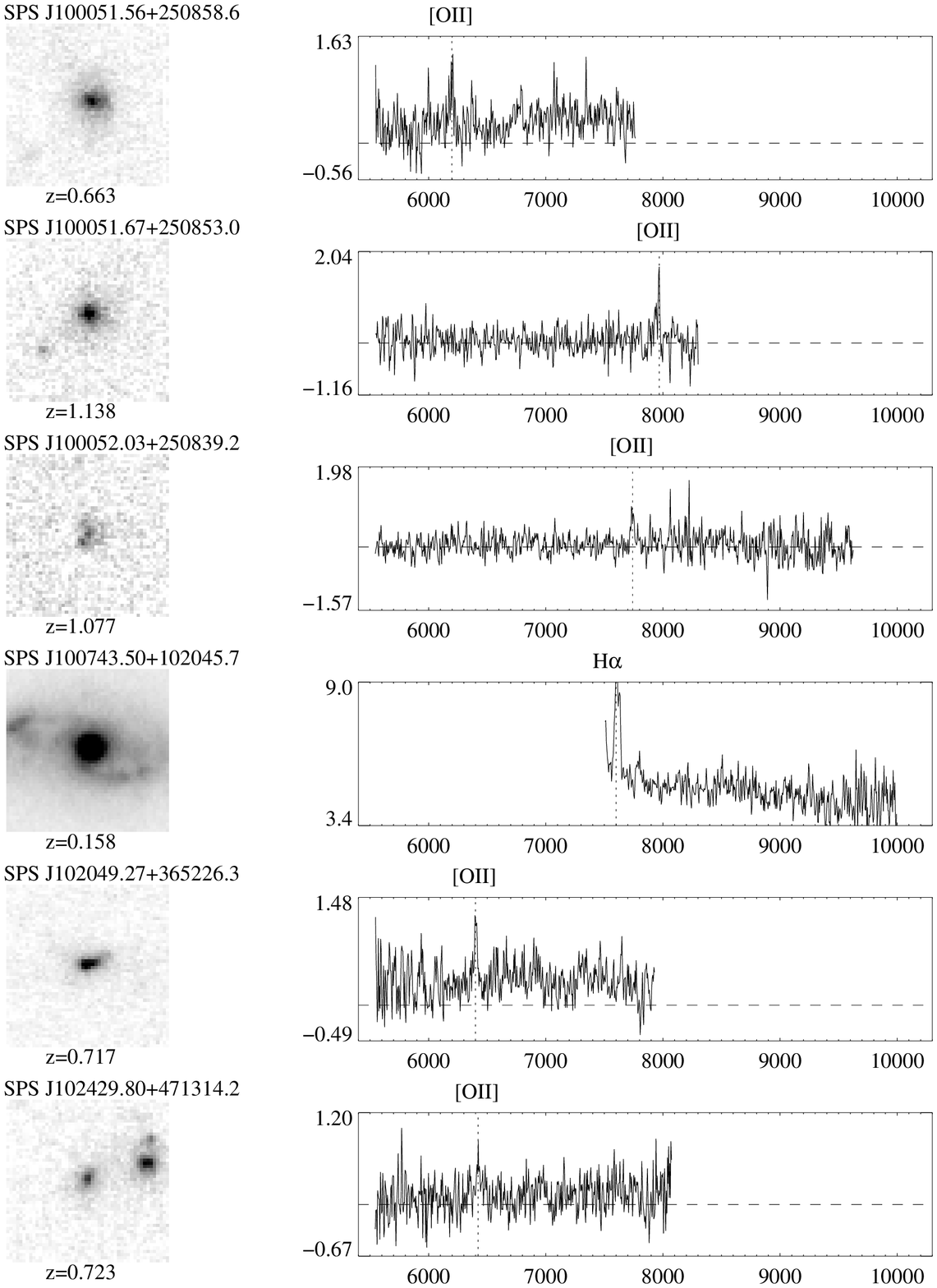}
\caption{The direct image and extracted spectrum for SPS emission line objects, as in Figure 3.}
\end{figure}

\clearpage
\begin{figure}
\plotone{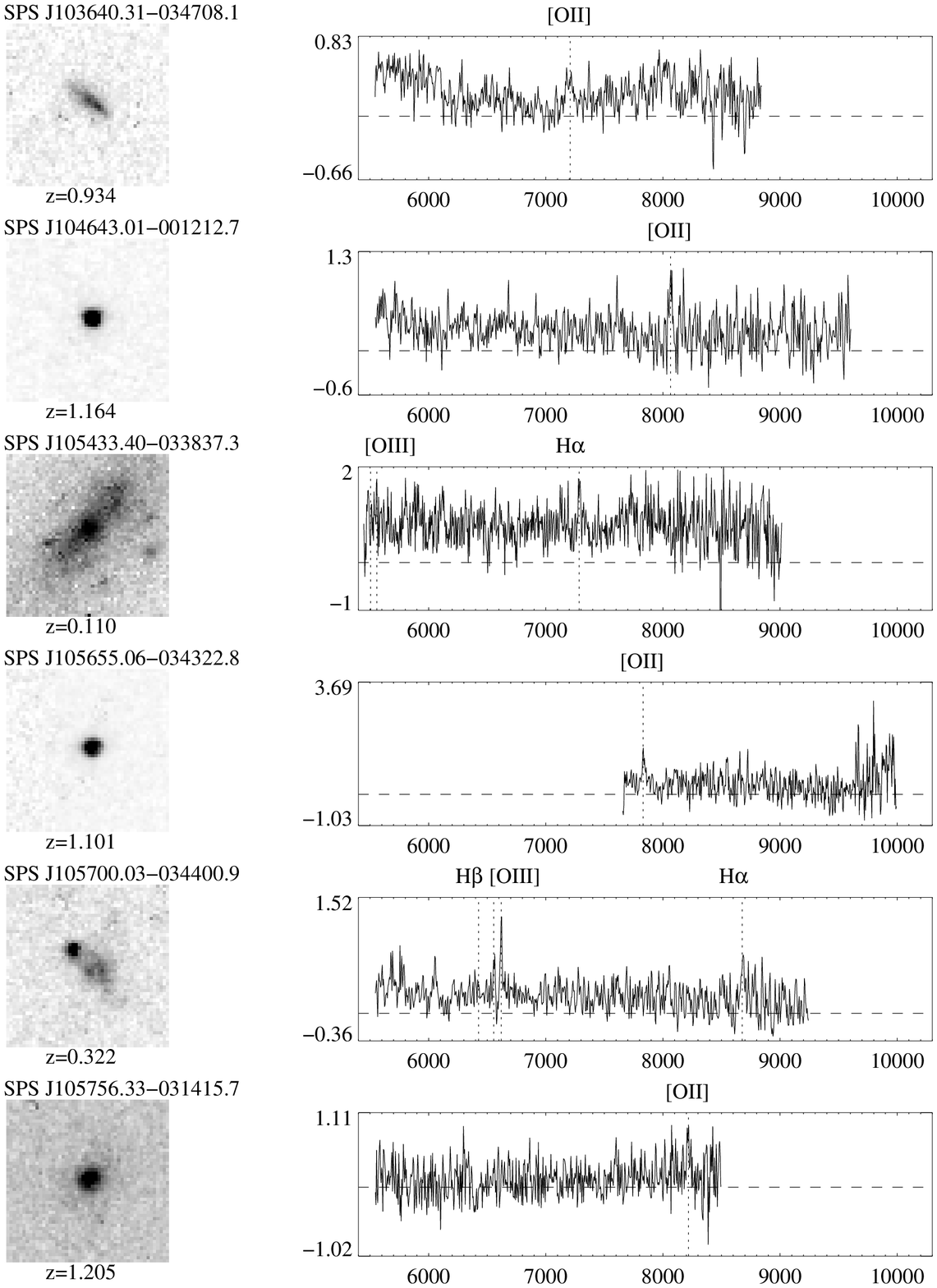}
\caption{The direct image and extracted spectrum for SPS emission line objects, as in Figure 3.}
\end{figure}

\clearpage
\begin{figure}
\plotone{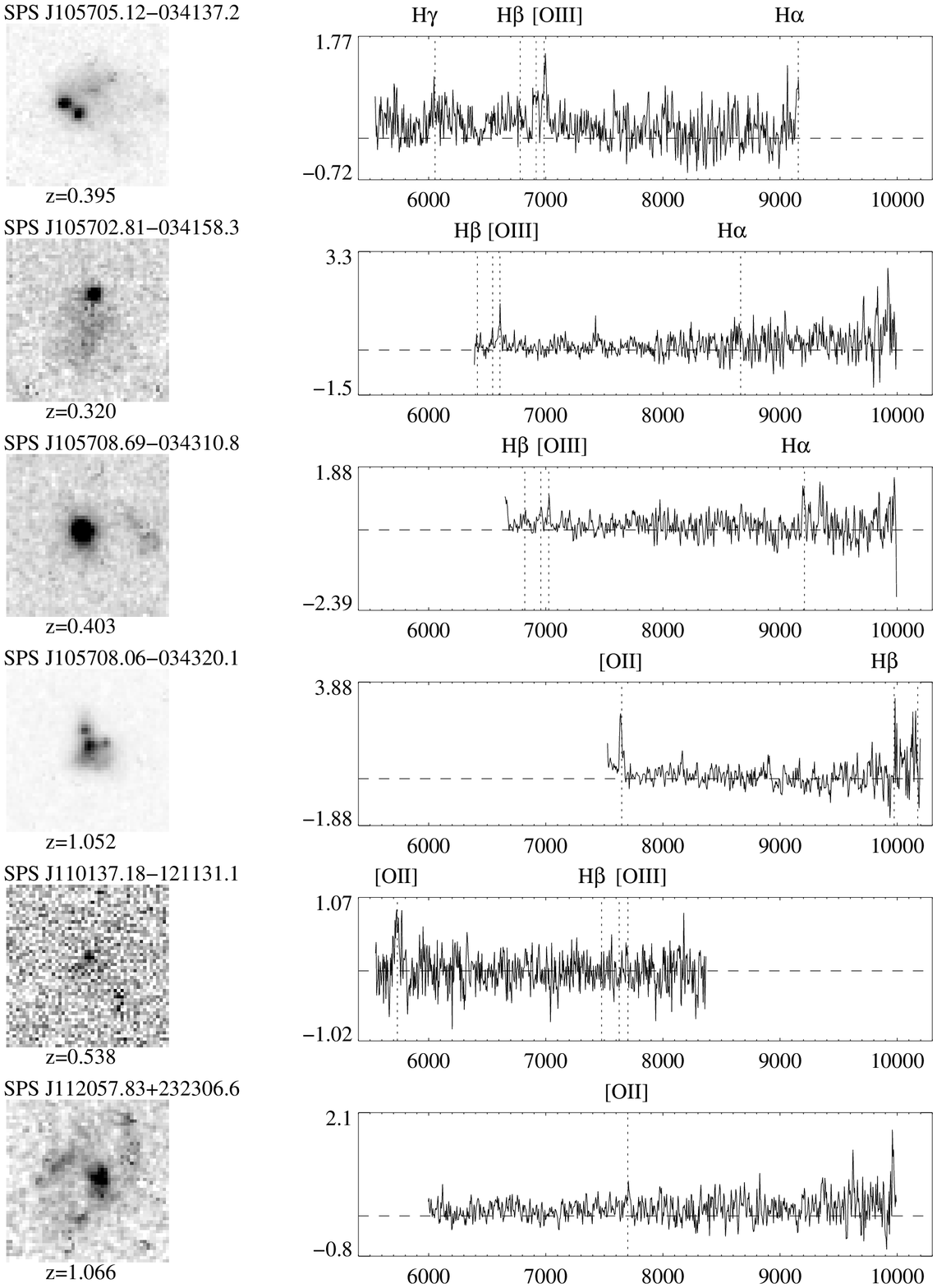}
\caption{The direct image and extracted spectrum for SPS emission line objects, as in Figure 3.}
\end{figure}

\clearpage
\begin{figure}
\plotone{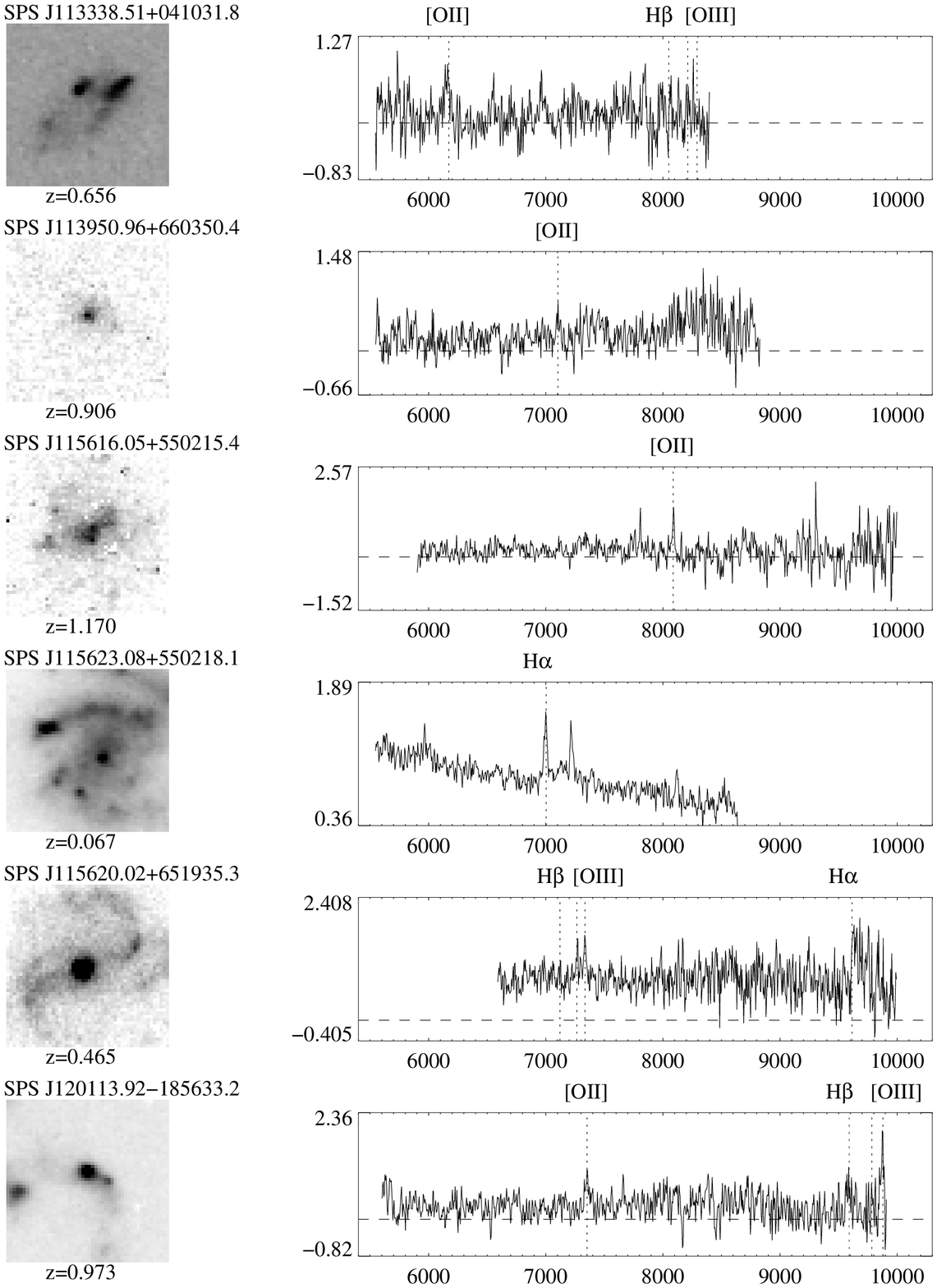}
\caption{The direct image and extracted spectrum for SPS emission line objects, as in Figure 3.}
\end{figure}

\clearpage
\begin{figure}
\plotone{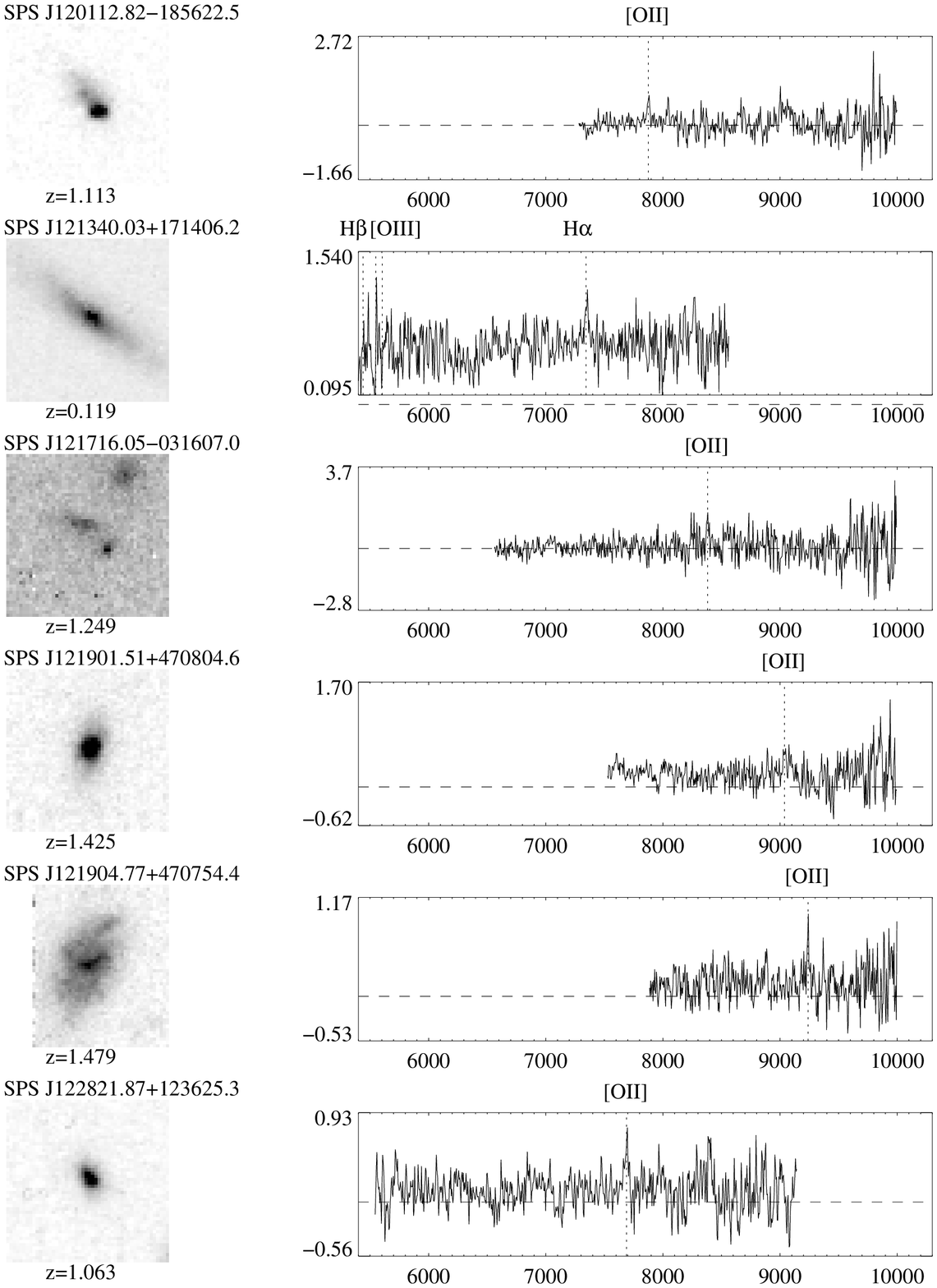}
\caption{The direct image and extracted spectrum for SPS emission line objects, as in Figure 3.}
\end{figure}

\clearpage
\begin{figure}
\plotone{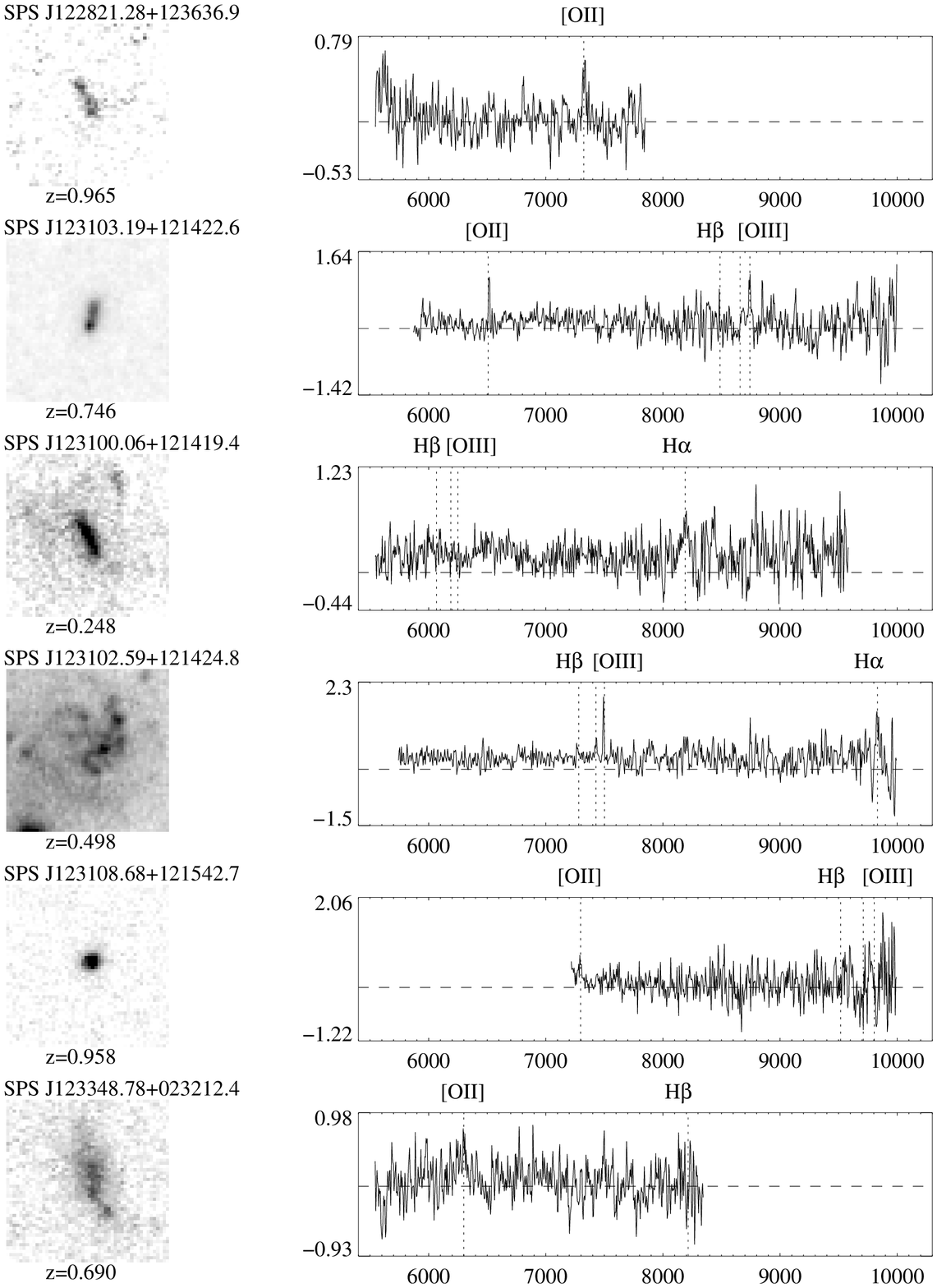}
\caption{The direct image and extracted spectrum for SPS emission line objects, as in Figure 3.}
\end{figure}

\clearpage
\begin{figure}
\plotone{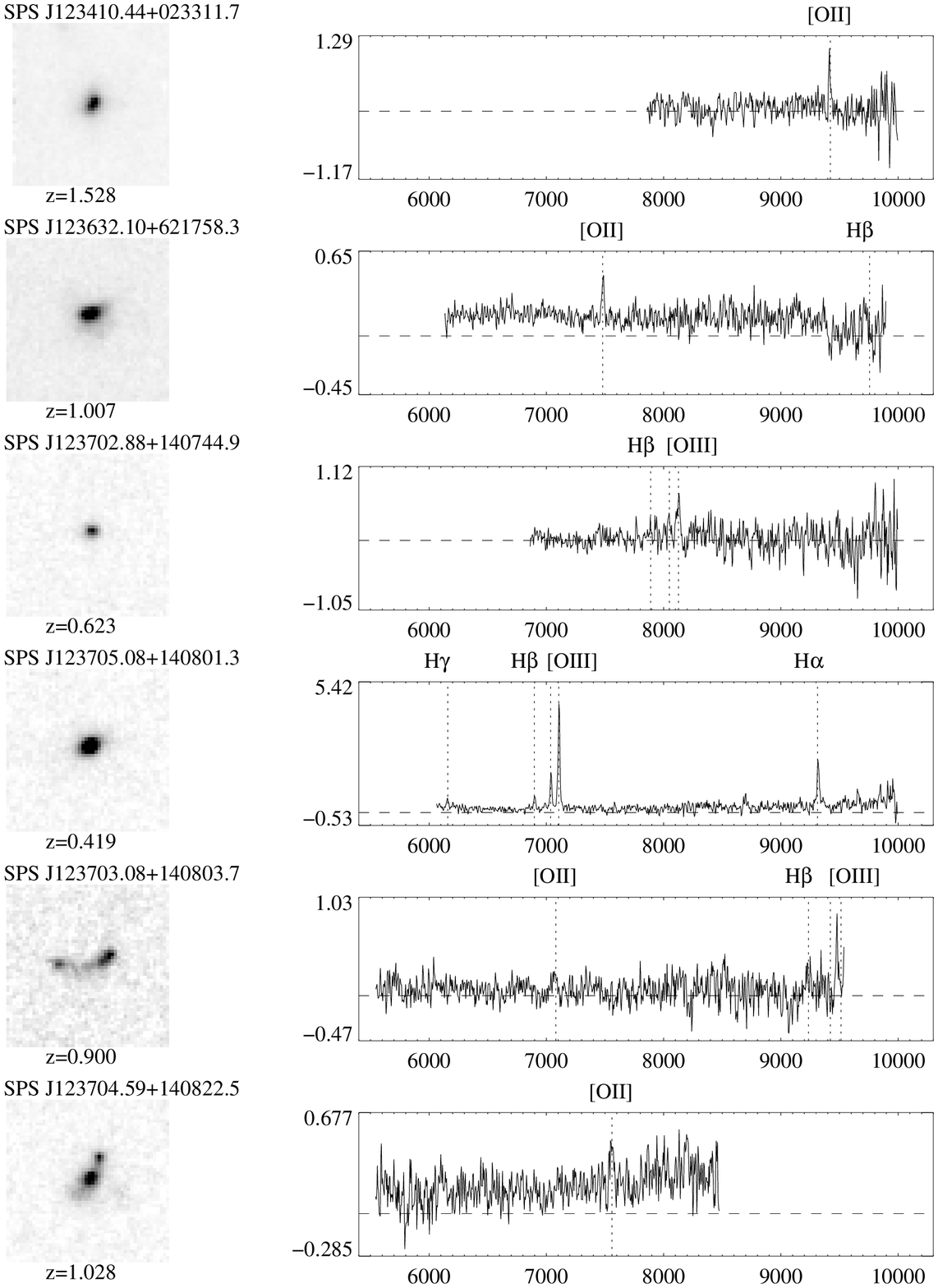}
\caption{The direct image and extracted spectrum for SPS emission line objects, as in Figure 3.}
\end{figure}

\clearpage
\begin{figure}
\plotone{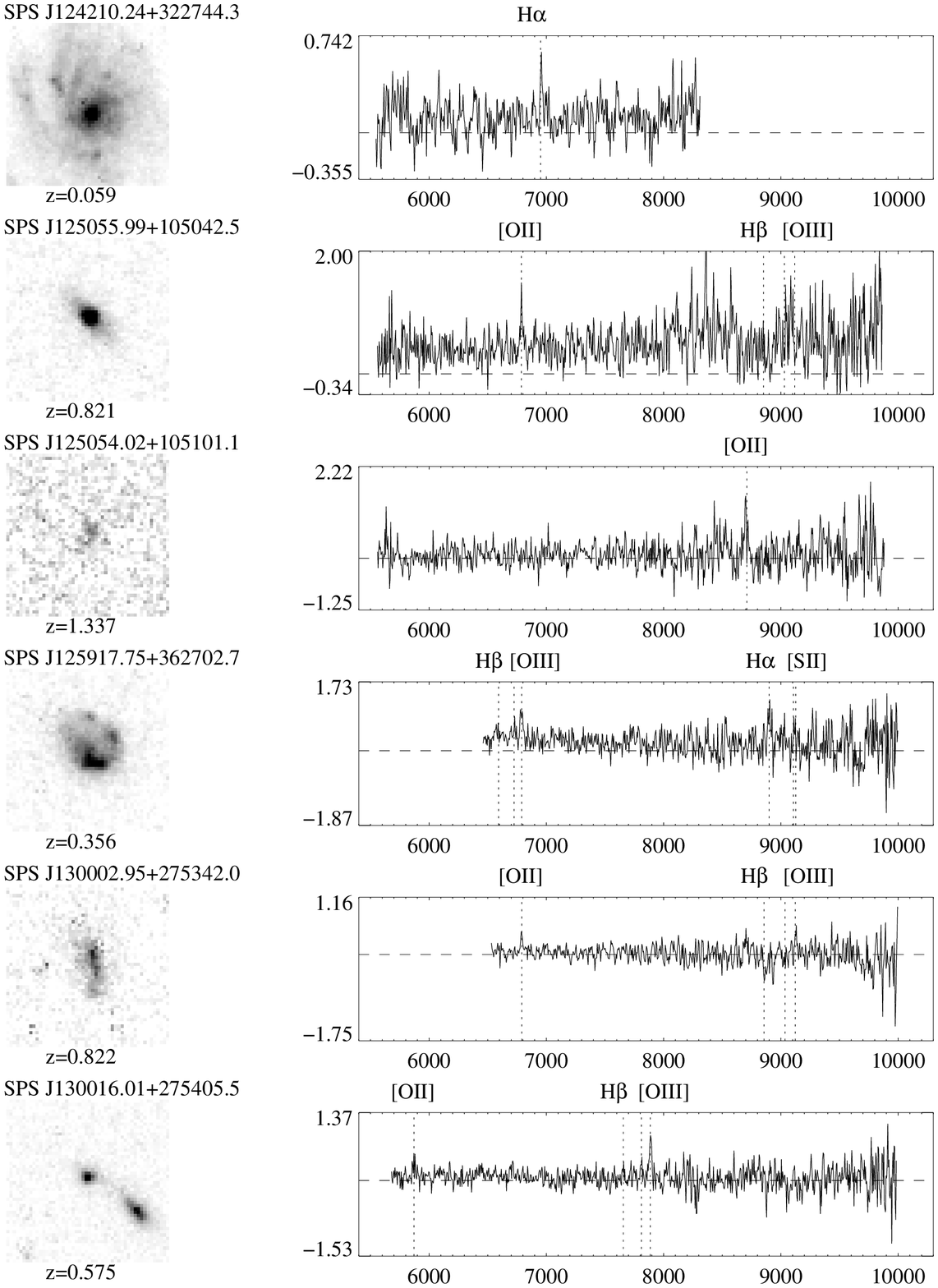}
\caption{The direct image and extracted spectrum for SPS emission line objects, as in Figure 3.}
\end{figure}

\clearpage
\begin{figure}
\plotone{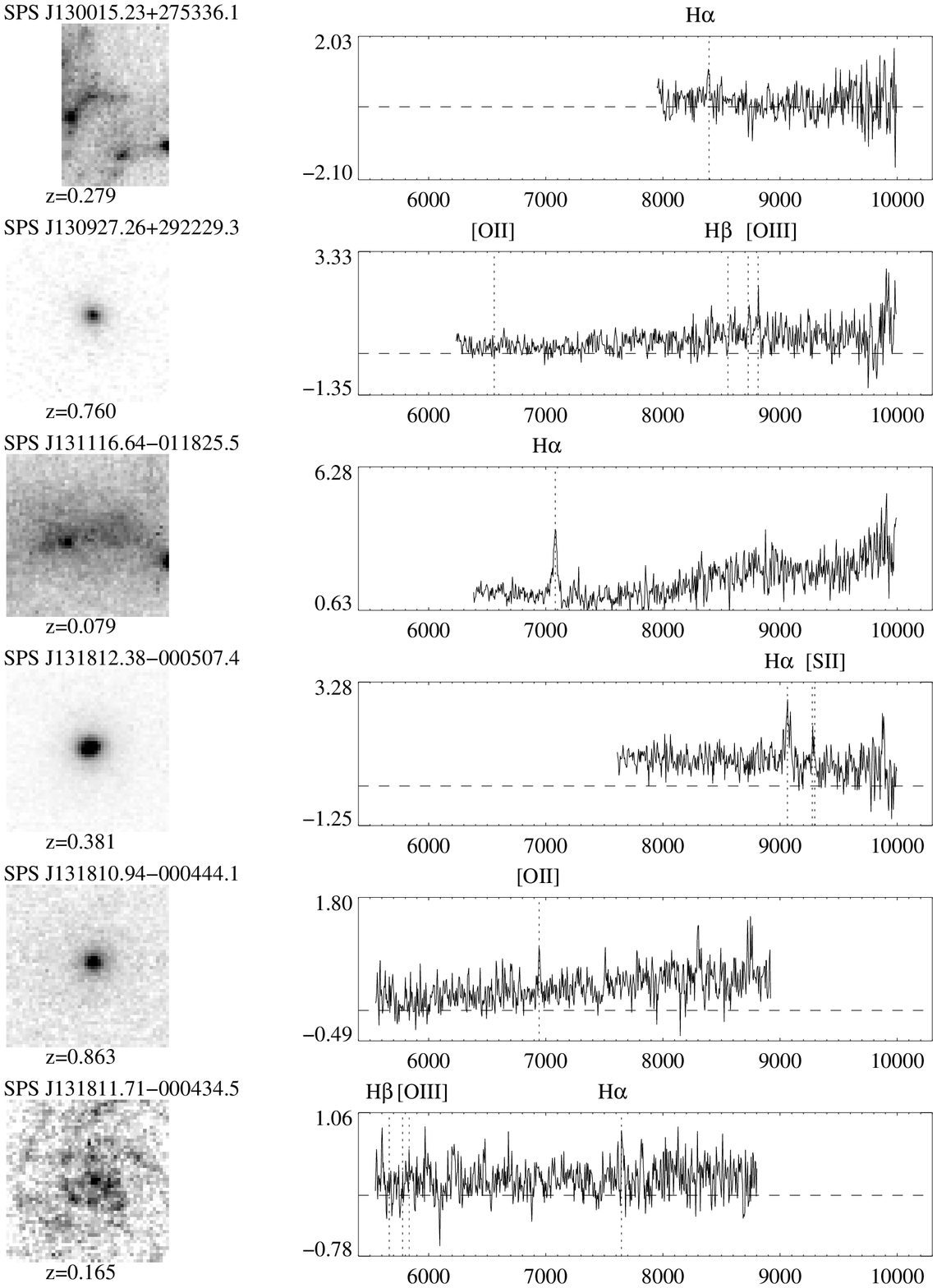}
\caption{The direct image and extracted spectrum for SPS emission line objects, as in Figure 3.}
\end{figure}

\clearpage
\begin{figure}
\plotone{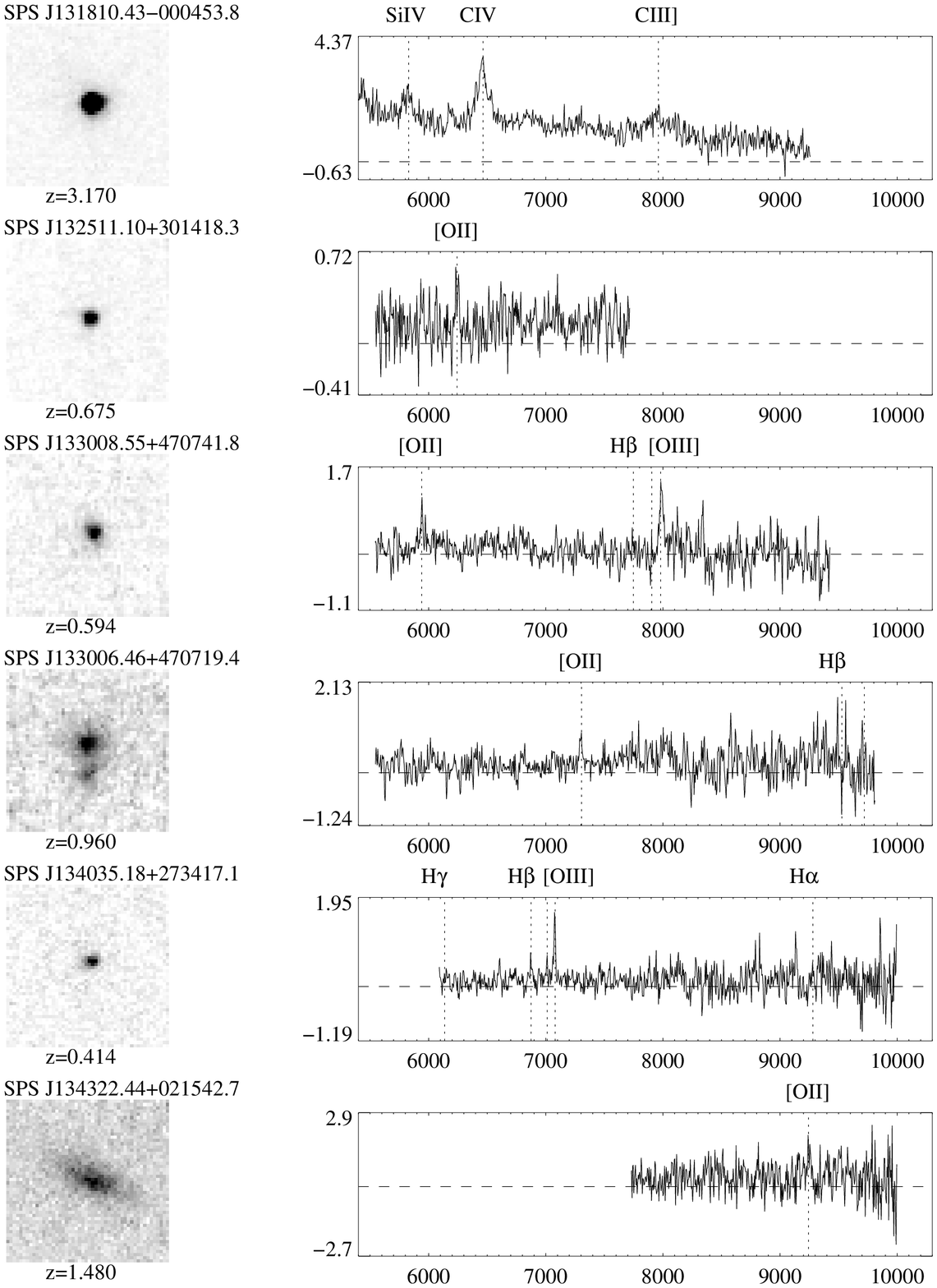}
\caption{The direct image and extracted spectrum for SPS emission line objects, as in Figure 3.}
\end{figure}

\clearpage
\begin{figure}
\plotone{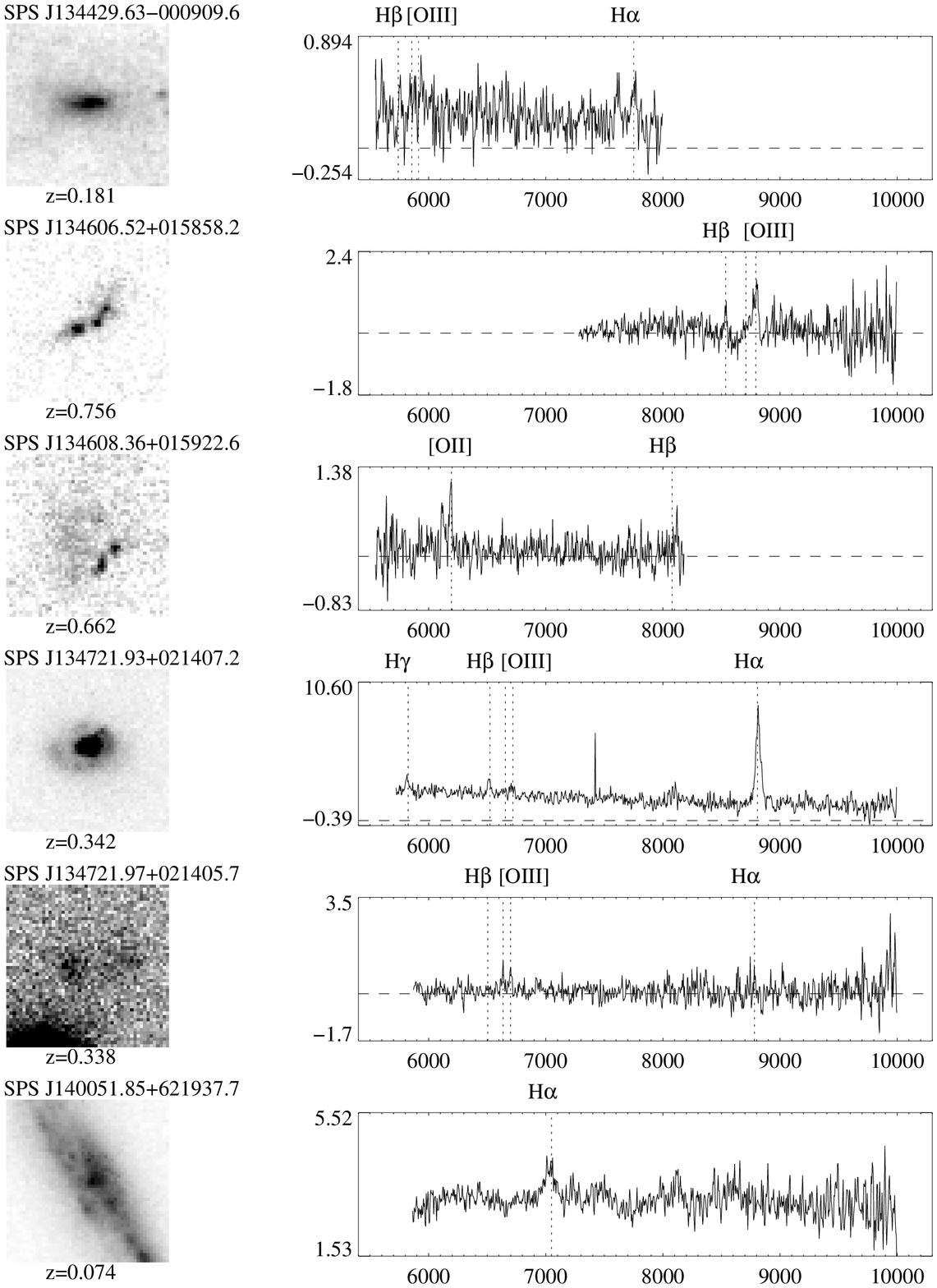}
\caption{The direct image and extracted spectrum for SPS emission line objects, as in Figure 3.}
\end{figure}

\clearpage
\begin{figure}
\plotone{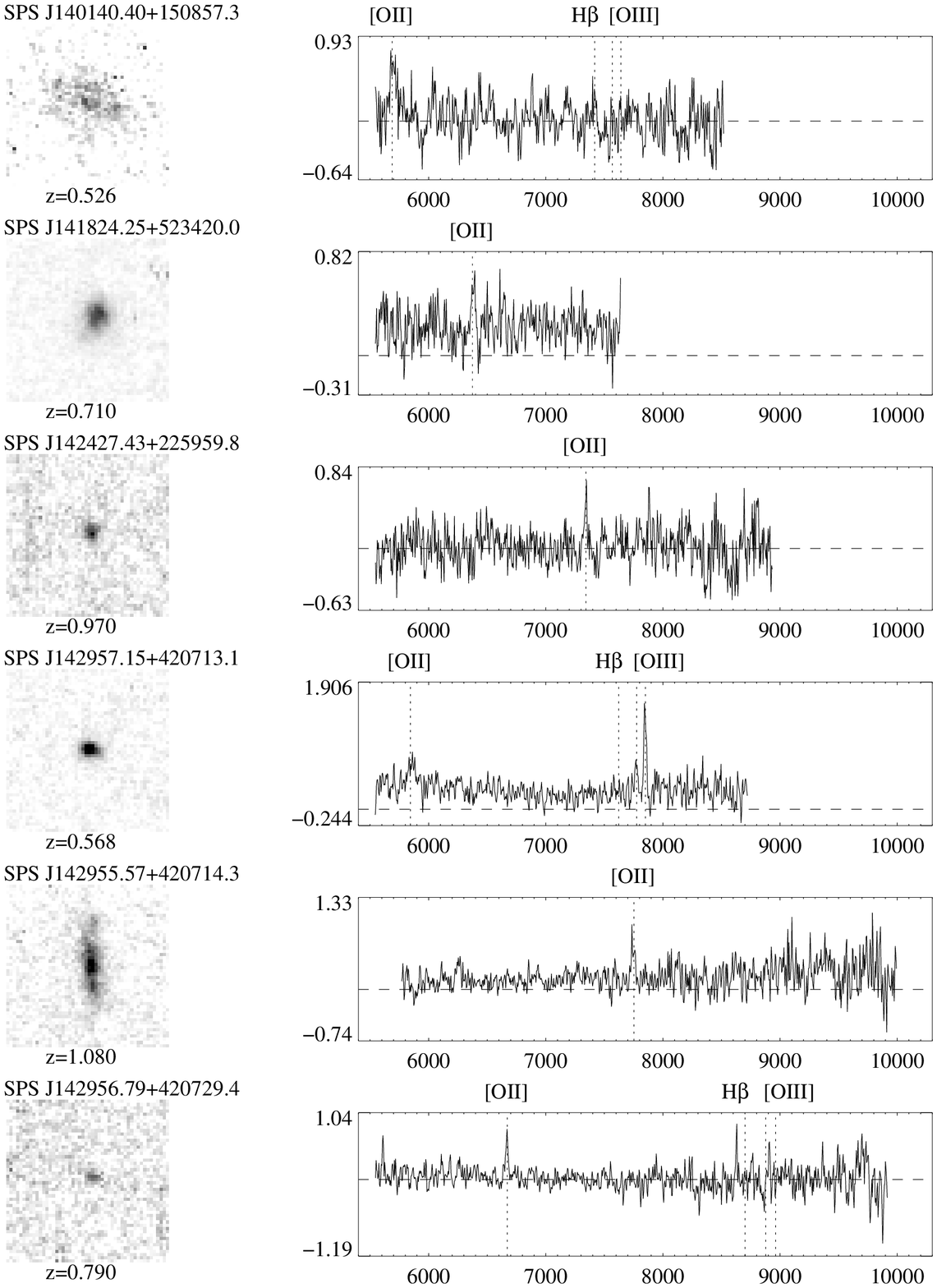}
\caption{The direct image and extracted spectrum for SPS emission line objects, as in Figure 3.}
\end{figure}

\clearpage
\begin{figure}
\plotone{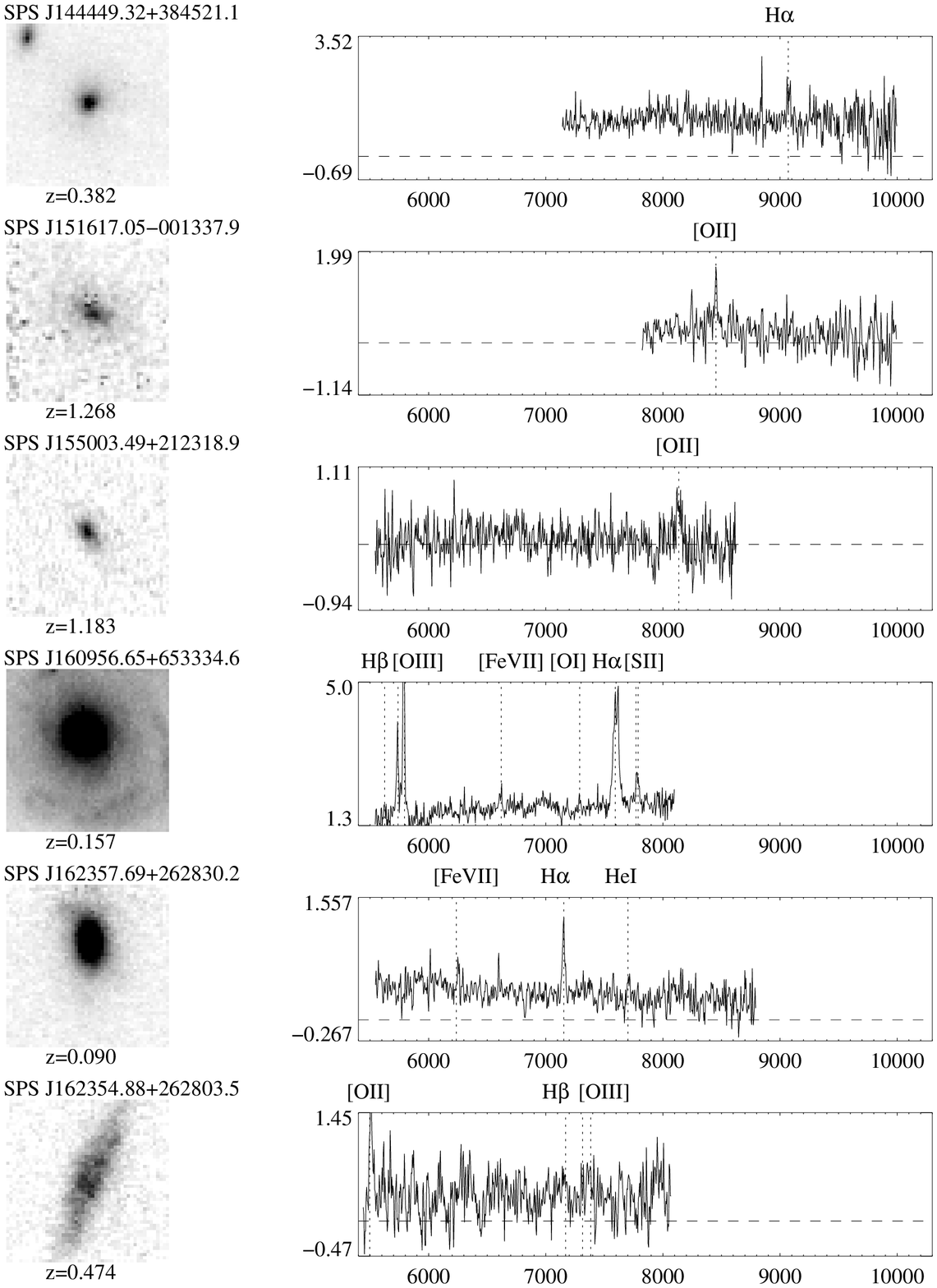}
\caption{The direct image and extracted spectrum for SPS emission line objects, as in Figure 3.}
\end{figure}

\clearpage
\begin{figure}
\plotone{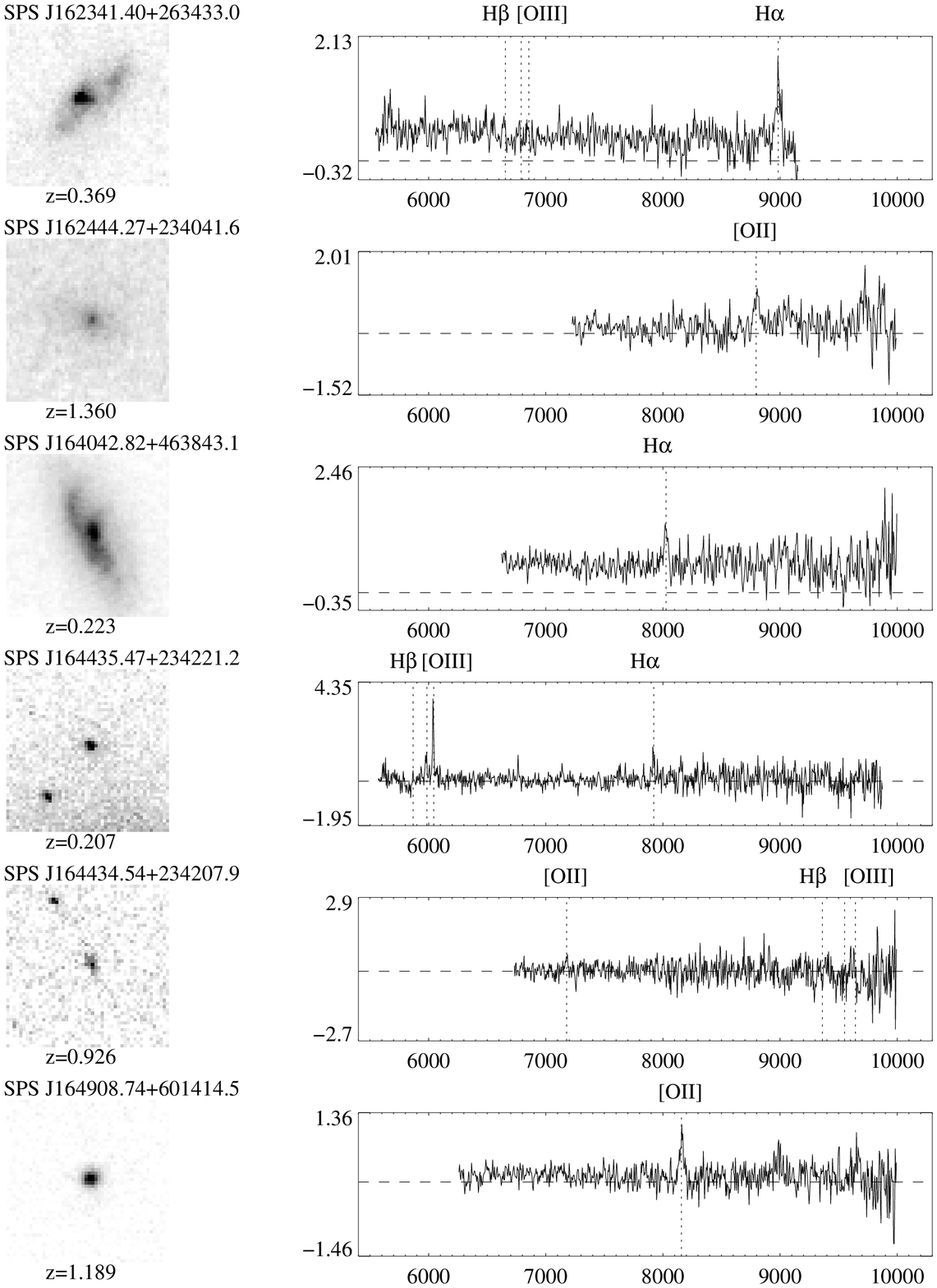}
\caption{The direct image and extracted spectrum for SPS emission line objects, as in Figure 3.}
\end{figure}

\clearpage
\begin{figure}
\plotone{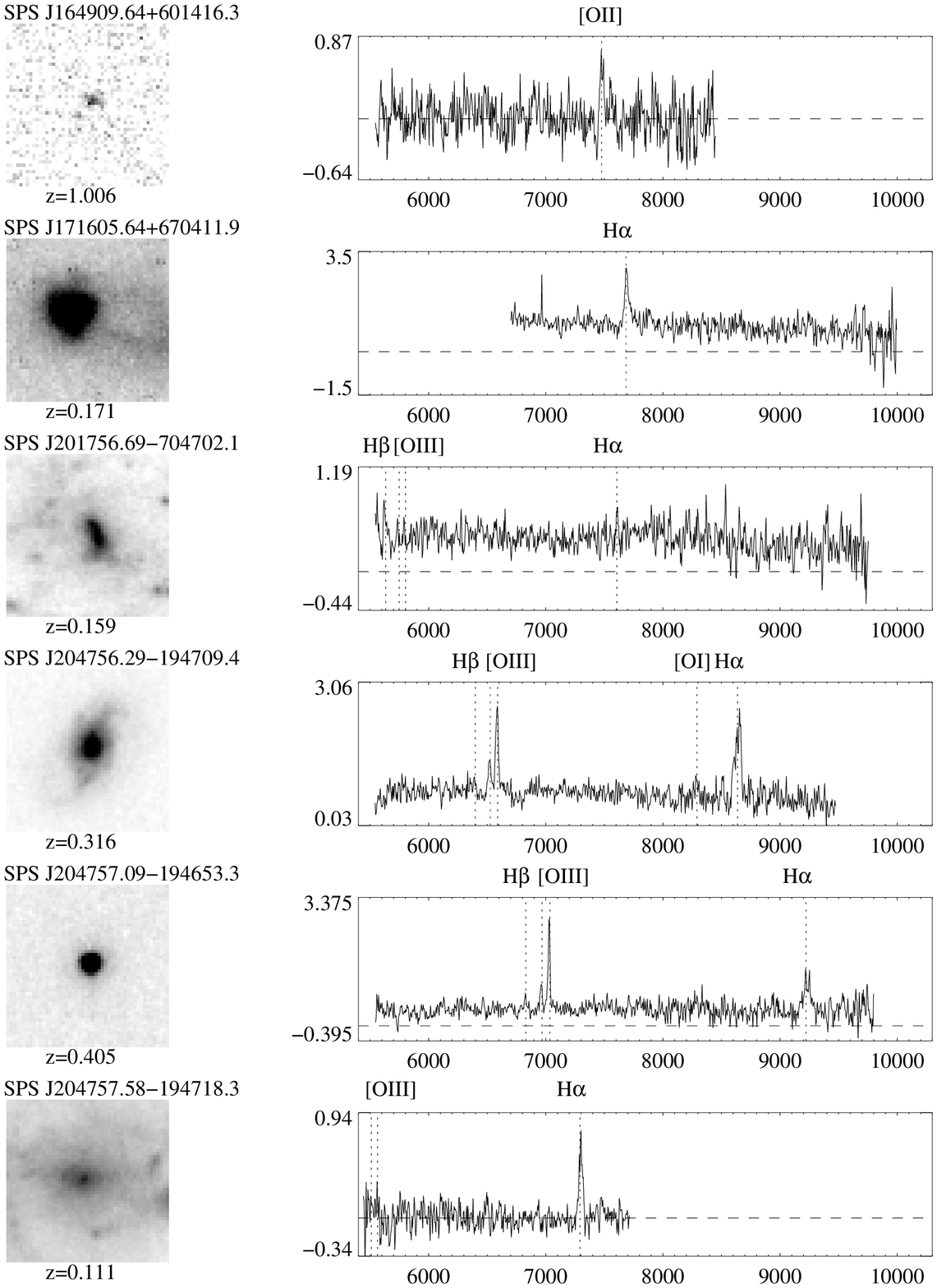}
\caption{The direct image and extracted spectrum for SPS emission line objects, as in Figure 3.}
\end{figure}

\clearpage
\begin{figure}
\plotone{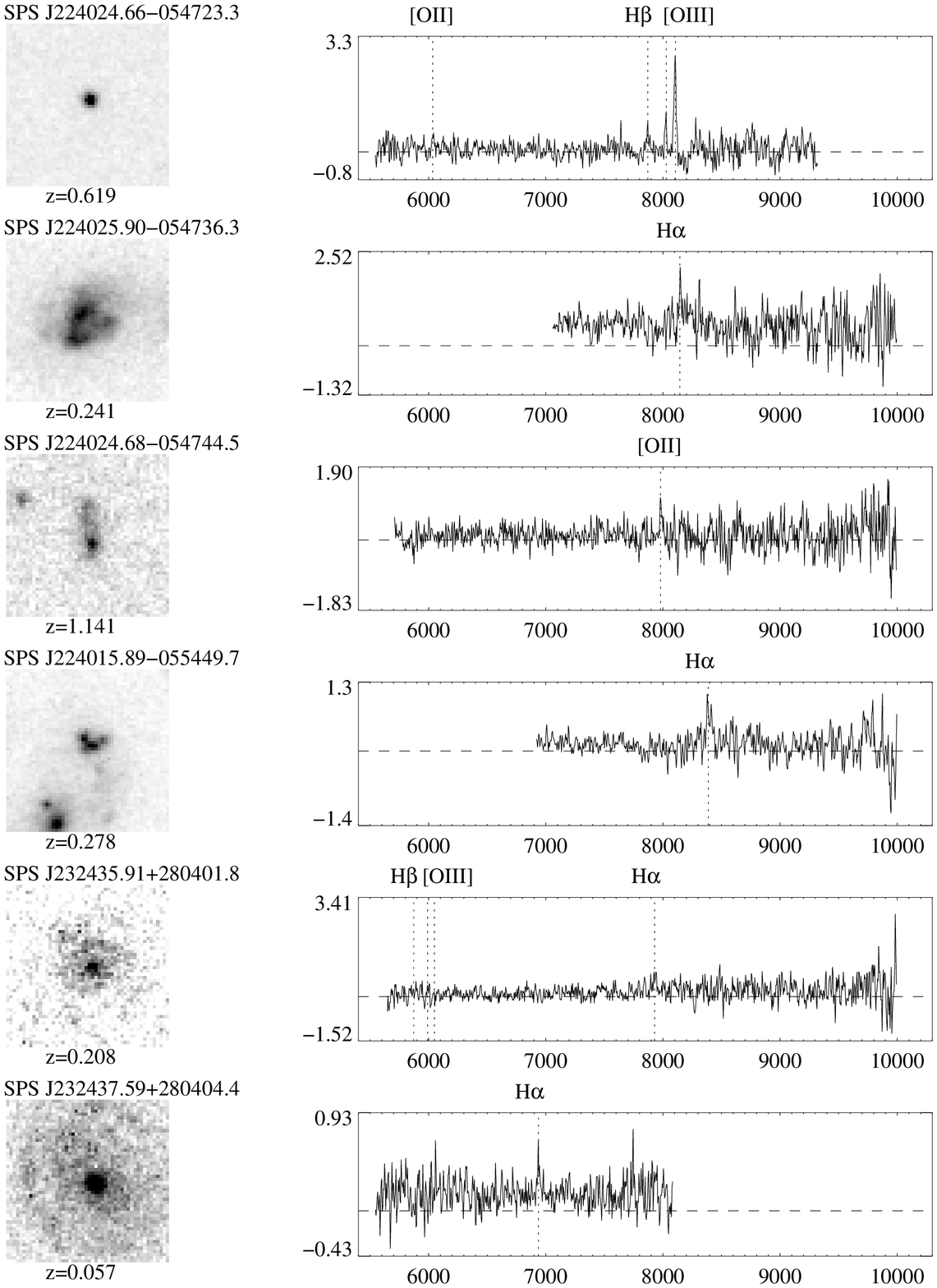}
\caption{The direct image and extracted spectrum for SPS emission line objects, as in Figure 3.}
\end{figure}

\clearpage
\begin{figure}
\plotone{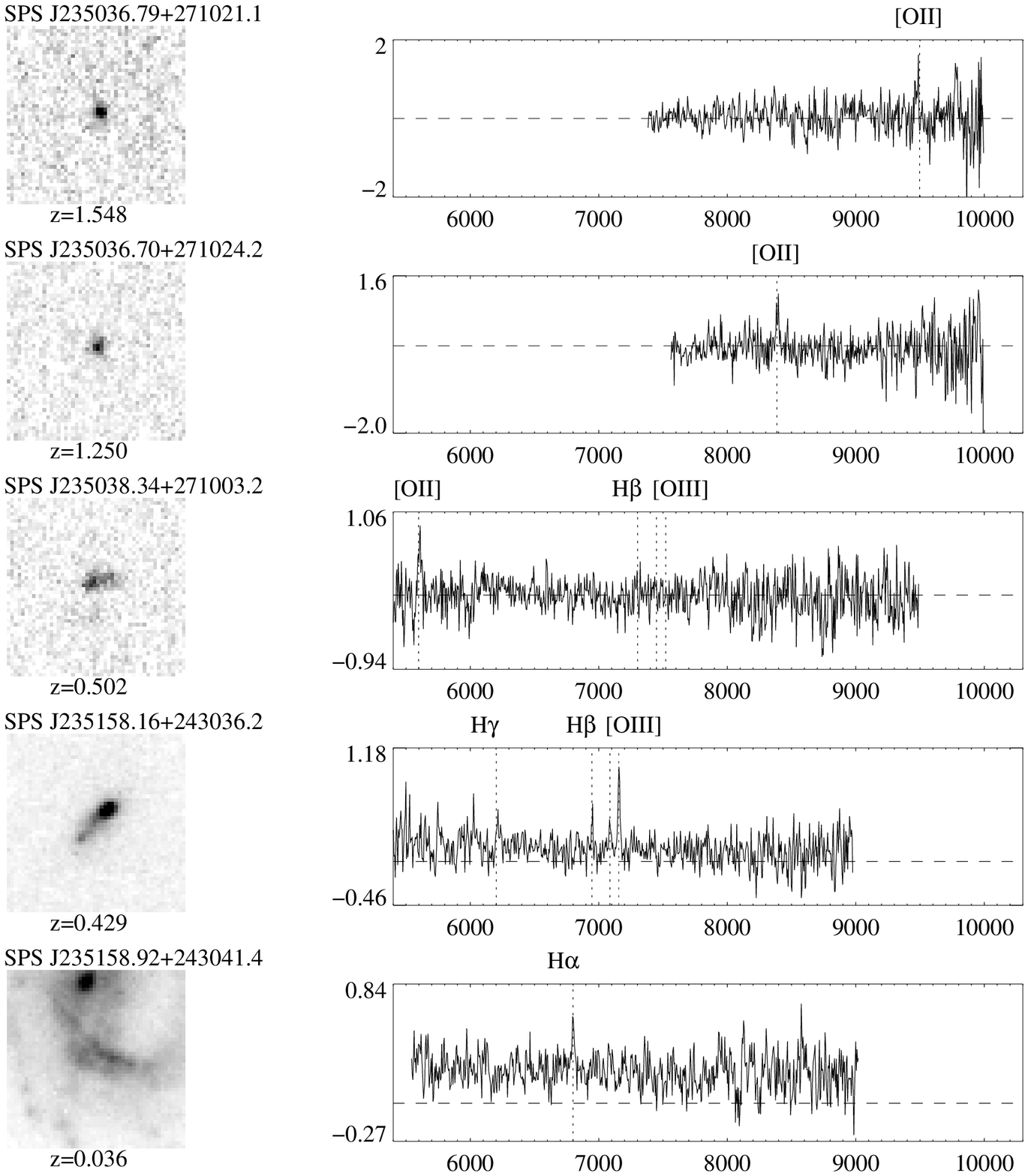}
\caption{The direct image and extracted spectrum for SPS emission line objects, as in Figure 3.}
\end{figure}
\clearpage

\begin{figure}
\plottwo{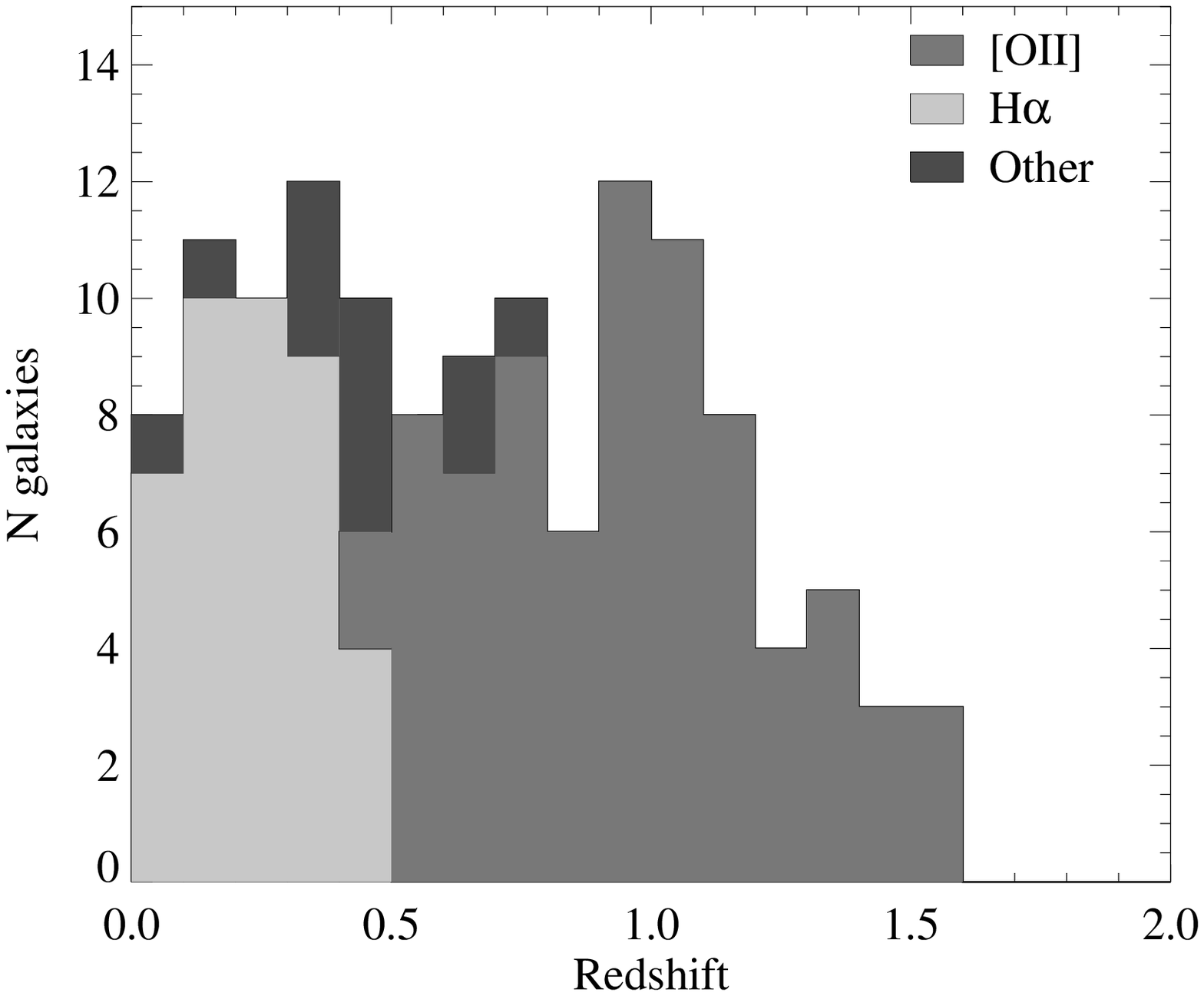}{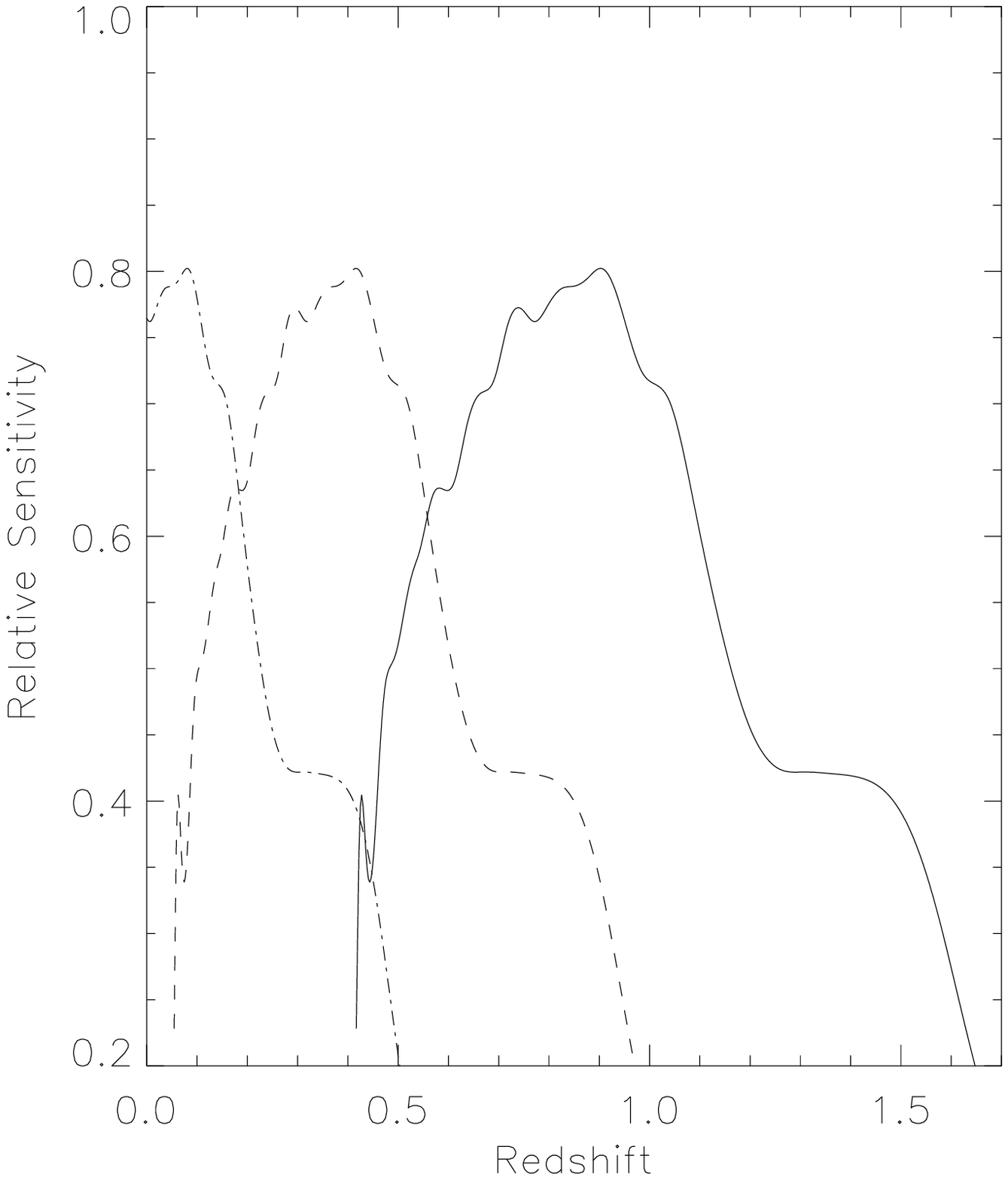}
\caption{\underline{a}:  The distribution of redshifts for emission line objects.  The medium grey
region indicates the redshift
of [OII] emitters, the light grey region that of \ha\ emitters, and the dark grey that objects with neither
[OII] nor \ha.  \underline{b}:  The relative sensitivity of G750L to each emission line as a function
of redshift.  The solid line indicates they sensitivity to redshift [OII] emission, the dashed line the
sensitivity to [OIII]5007 emission, and the dot-dashed line the sensitivity to \ha\ emission.
\label{fig: zhist} }
\end{figure}
\clearpage

\begin{figure}
\plotone{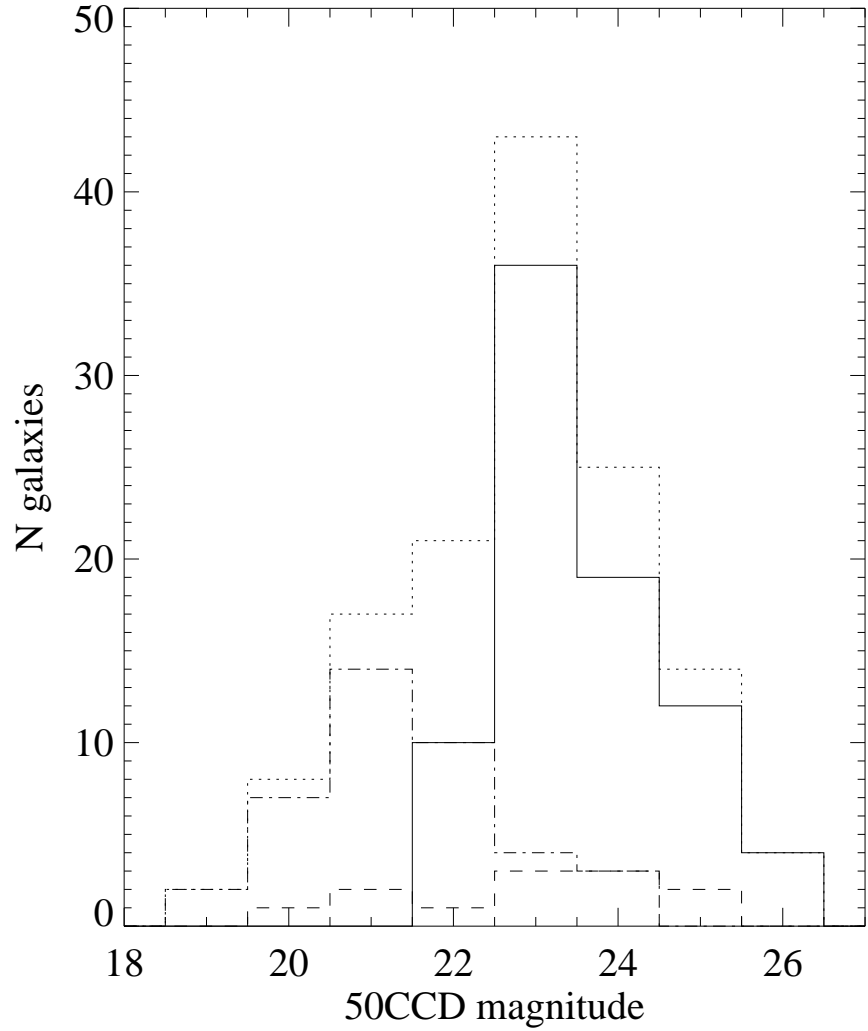}
\caption{The histogram of object photometry for emission line objects.  The dotted line shows the histogram
for all emission line objects.  The solid line indicates the magnitude
of [OII] emitters, the dot-dashed line that of \ha\ emitters and the dashed line that objects with neither
[OII] nor \ha.  
\label{fig: maghist} }
\end{figure}

\end{document}